\documentclass[11pt,a4paper]{article}

\usepackage{amssymb}
\usepackage{amsmath}
\usepackage[pdftex]{graphicx}
\usepackage{graphics}
\usepackage{epsfig}
\usepackage{color}
\usepackage{media9}
\usepackage[colorlinks]{hyperref}
\usepackage{listings}
\hypersetup{linkcolor=blue,%
	citecolor=red,%
	urlcolor=cyan}
\usepackage{url} 

\newcommand{\be}{\begin{equation}}
\newcommand{\ee}{\end{equation}}
\numberwithin{equation}{section}

\title{\bf RBF Solver for Quaternions Interpolation}

\author{F. Rinaldi$^{1,2}$\footnote{f.rinaldi@unimarconi.it, ORCID:  \href{http://orcid.org/0000-0002-0087-3042}{0000-0002-0087-3042}}
	, D. Dolci $^3$\footnote{danieledolci95@gmail.com}
	\\ [2ex]
	\footnotesize \sl $^1$Dipartimento di Fisica Nucleare, Subnucleare e delle Radiazioni, \\ 
	\footnotesize \sl Univ. degli Studi Guglielmo Marconi,Via Plinio 44, I-00193 Rome, Italy\\
	\footnotesize \sl $^2$CERFIM, PO Box 1132, CH-6601 Locarno, Switzerland\\	
	\footnotesize \sl $^3$Technical Director and Rigging Instructor, Department of Rigging,\\
	\footnotesize \sl Animschool, 209 West 520 North, Orem, UT, 84057, USA\\
}

\begin{document}
	
	\maketitle

\begin{abstract}
	In this paper we adapt the RBF Solver to work with quaternions by taking advantage of their Lie Algebra and exponential map. This will allow to work with quaternions as if they were normal vectors in ${{\mathbb R}}^3$ and blend them in a very efficient way.
\end{abstract}

\bigskip 

\section{Introduction}

Transforming objects in the space is quite a tough challenge, especially when it comes down to rotations. These operations are essential in computer graphics and over the last few decades many algorithms and mathematical models have been developed.
One of the most common problems encountered while dealing with rotations is the gimbal lock, which is generally overcome by the usage of Quaternions. When Hamilton firstly discovered Quaternions \cite{Ham}, he had no idea they would relate to Galois Theorie \cite{Mil}, which was only proved in 1981 \cite{Dea}, or that quaternions would be able to generate a Lie algebra, which really comes handy to create a locally linearized map to operate on.
This last point lays the groundwork for this paper which provides a possible way of creating an RBF solver (nowadays widely used in computer graphics), by blending Quaternions (using their Lie Algebra) rather than ordinary vectors.
\\
The CG (Computer Graphics) is one of those fields where Quaternions become particularly successful, see for example \cite{Sho}. This is also true for the RBF Solver, which over the last few years has found a lot of applications in this field. The idea behind this paper is to blend multiple Quaternions rather than vectors with this solver. In order to do so it will be shortly presented some notion of transformations, $Gimbal$ $Lock$ and Quaternions Interpolation. The paper also contains an example of how this solver could be used in production.

\subsection{\textit{Describing transformations.}}

It is well known that that each and all the isometries in ${{\mathbb R}}^n$ can be written as:\begin{equation}
f{\rm :}{{\mathbb R}}^n{\rm \ }\to {{\mathbb R}}^n,   f\left(x\right){\rm =}Mx{\rm +}b,
\end{equation}
\\
where $M$ is an orthogonal matrix and $b$ is a vector in ${{\mathbb R}}^n$.  
\\
A crucial theorem states that every linear transformation is the composition of a linear isometry, a dilation and a shear. Algebraically, this means that every invertible matrix $M$ can be written as the product 
\\
\begin{equation}
M{\rm =}KAN,
K{\rm =}\left( \begin{array}{cc}
k_{{\rm 1}} & k_{{\rm 2}} \\ 
k_{{\rm 3}} & k_{{\rm 4}} \end{array}
\right), A{\rm =}\left( \begin{array}{cc}
\lambda  & 0 \\ 
0 & \mu  \end{array}
\right), 
N{\rm =}\left( \begin{array}{cc}
{\rm 1} & a \\ 
0 & {\rm 1} \end{array}
\right).
\end{equation}

\begin{equation}
M{\rm =}\left( \begin{array}{cc}
m_{{\rm 1}} & m_{{\rm 2}} \\ 
m_{{\rm 3}} & m_{{\rm 4}} \end{array}
\right){\rm =}\left( \begin{array}{cc}
k_{{\rm 1}} & k_{{\rm 2}} \\ 
k_{{\rm 3}} & k_{{\rm 4}} \end{array}
\right)\left( \begin{array}{cc}
\lambda  & 0 \\ 
0 & \mu  \end{array}
\right)\left( \begin{array}{cc}
{\rm 1} & a \\ 
0 & {\rm 1} \end{array}
\right)
\end{equation}
\\
Translation, Scaling, homothety, reflection, rotation, shearing etc are all considered affine transformations, meaning a transformation between affine spaces that preserves point, straight lines and planes.  All of those transformations, together, form the group of affine transformations of whose  $E{\rm (}n{\rm )}$ is a subgroup. 
\\
As stated before the isometries in $E{\rm (}n{\rm )}$ can be formed by a translation together with an orthogonal (linear) transformation. In terms of abstract algebra, this means that $E{\rm (}n{\rm )}$ can be splitted into 2 subgroups, one that contains the translations, denoted $T{\rm (}n{\rm )}$ and called translational group, the other, denoted $O\left(n\right)$, is the orthogonal group. 
\\
The relation between those is:
\begin{equation}
O\left(n\right)\cong E{\rm (}n{\rm )/}T{\rm (}n{\rm )}
\end{equation}
\\
And therefore $O{\rm (}n{\rm )}$ is the quotient group of the mod operation between $E\left(n\right)$ and $T{\rm (}n{\rm )}$. $O{\rm (}n{\rm )}$ has a subgroup called special orthogonal group and denoted $SO{\rm (}n{\rm )}$.
\\
We know that $E{\rm (}n{\rm )}$ has a subgroup of direct isometries which are those formed by the combination of a translation with one of the isometries contained in $SO{\rm (}n{\rm )}$. This is usually formalized as:
\begin{equation}
SO\left(n\right)\cong E^{{\rm +}}{\rm (}n{\rm )/}T{\rm (}n{\rm )}
\end{equation}
\\
This insight of the subgroups of $E{\rm (}n{\rm )}$ it will be helpful in understanding the $Gimbal$ $Lock$.

\subsection{\textit{Gimbal Lock}}

The term $Gimbal$ $Lock$ has been used improperly to refer to a mathematical singularity. The reason why its use is considered improper is that no real lock happens. Indeed, in its most generic meaning, the $Gimbal$ $Lock$ is the loss of rotation freedom about one of the axis when a certain angle is hit by another axis. 
\\
In order to fully understand why $Gimbal$ $Lock$ occurs it is necessary to dive into the topology of the rotation group ${\rm SO(3)}$ which will be done in Section 2. 

\subsection{\textit{ What is an RBF solver?}}

This kind of solver allows to use a certain amount of positions to interpolate between the same amount of samples. 
\\
A position can either be a physical position in the space described by a transformation matrix, or something purely abstract. For example we could define happiness, sadness and angriness as three different positions, we could say that those are described by a matrix of values of the current state of the mouth corners, eyebrows, upper and lower lids. The values describing the position are gathered in a N-dimensional key. This key, in the case of a transformation matrix (${\rm 3\times 3}$ matrix), could be written as a ${\rm 1\times 9}$ matrix.  In the scenario of the facial expression, the key describing the position would be a ${\rm 1\times 4}$ matrix. 
\\
To each of those positions we associate a sample, which is a M-dimensional matrix. The power of this solver is that you can have $K$ amount of N-dimensional key in input driving $K$ amount M-dimensional output (by blending the samples). 

\section{Inside the Gimbal Lock}

\subsection{\textit{Matrix representation}}

When trying to rotate the point ${\rm P}$ about more than one axis what is really happening is the composition of rotations one after the other. It is important to remember that rotation composition $is$ $not$ commutative.  In this sense it is possible to visualize the 3 arbitrary axis in a hierarchical structure where they are parented and dependent to each other according to a given order.
\\
\noindent
\textbf{Example.} To demonstrate this structure it is possible to simply use matrix product which is also not commutative. In particular, let's imagine a point ${\rm P(1,\ 1,\ 1)\ }$in a Right-Hand Cartesian System that needs to be rotate $\frac{\pi }{{\rm 2}}$  about each axis. 
\\
The rotation matrices will therefore look as follow:

\begin{align}
R_x\left(\varphi \right)P{\rm =}\left[ \begin{array}{ccc}
{\rm 1} & 0 & 0 \\ 
0 & {\cos  \varphi \ } & {\rm -}{\sin  \varphi \ } \\ 
0 & {\sin  \varphi \ } & {\cos  \varphi \ } \end{array}
\right]\left[ \begin{array}{c}
P_x \\ 
P_y \\ 
P_z \end{array}
\right],\\
R_y\left(\varphi \right)P{\rm =}\left[ \begin{array}{ccc}
{\cos  \varphi \ } & 0 & {\sin  \varphi \ } \\ 
0 & {\rm 1} & 0 \\ 
{\rm -}{\sin  \varphi \ } & 0 & {\cos  \varphi \ } \end{array}
\right]\left[ \begin{array}{c}
P_x \\ 
P_y \\ 
P_z \end{array}
\right],\\
R_z\left(\varphi \right)P{\rm =}\left[ \begin{array}{ccc}
{\cos  \varphi \ } & {\rm -}{\sin  \varphi \ } & 0 \\ 
{\sin  \varphi \ } & {\cos  \varphi \ } & 0 \\ 
0 & 0 & {\rm 1} \end{array}
\right]\left[ \begin{array}{c}
P_x \\ 
P_y \\ 
P_z \end{array}
\right],
\end{align}
\\
where $\varphi =\frac{\pi }{{\rm 2}}$.
\\
\\
\noindent
After having defined a rotation order, which in this example will be  $x\to y\to z$, and doing some simple calculations, the composition of rotations lead to this result:

\begin{equation}
R_x\left(\frac{\pi }{{\rm 2}}\right)\left[ \begin{array}{c}
{\rm 1} \\ 
{\rm 1} \\ 
{\rm 1} \end{array}
\right]{\rm =}\left[ \begin{array}{ccc}
{\rm 1} & 0 & 0 \\ 
0 & {\cos  \frac{\pi }{{\rm 2}}\ } & {\rm -}{\sin  \frac{\pi }{{\rm 2}}\ } \\ 
0 & {\sin  \frac{\pi }{{\rm 2}}\ } & {\cos  \frac{\pi }{{\rm 2}}\ } \end{array}
\right]\left[ \begin{array}{c}
{\rm 1} \\ 
{\rm 1} \\ 
{\rm 1} \end{array}
\right]{\rm =}\left[ \begin{array}{ccc}
{\rm 1} & 0 & 0 \\ 
0 & 0 & {\rm -}{\rm 1} \\ 
0 & {\rm 1} & 0 \end{array}
\right]\left[ \begin{array}{c}
{\rm 1} \\ 
{\rm 1} \\ 
{\rm 1} \end{array}
\right]{\rm =}\left[ \begin{array}{c}
{\rm 1} \\ 
{\rm -}{\rm 1} \\ 
{\rm 1} \end{array}
\right],
\end{equation}

\begin{equation}
R_y\left(\frac{\pi }{{\rm 2}}\right)\left[{\rm -} \begin{array}{c}
{\rm 1} \\ 
{\rm 1} \\ 
{\rm 1} \end{array}
\right]{\rm =}\left[ \begin{array}{ccc}
{\cos  \frac{\pi }{{\rm 2}}\ } & 0 & {\sin  \frac{\pi }{{\rm 2}}\ } \\ 
0 & {\rm 1} & 0 \\ 
{\rm -}{\sin  \frac{\pi }{{\rm 2}}\ } & 0 & {\cos  \frac{\pi }{{\rm 2}}\ } \end{array}
\right]\left[ \begin{array}{c}
{\rm 1} \\ 
{\rm -}{\rm 1} \\ 
{\rm 1} \end{array}
\right]{\rm =}\left[ \begin{array}{ccc}
0 & 0 & {\rm 1} \\ 
0 & {\rm 1} & 0 \\ 
{\rm -}{\rm 1} & 0 & 0 \end{array}
\right]\left[ \begin{array}{c}
{\rm 1} \\ 
{\rm -}{\rm 1} \\ 
{\rm 1} \end{array}
\right]{\rm =}\left[ \begin{array}{c}
{\rm 1} \\ 
{\rm -}{\rm 1} \\ 
{\rm -}{\rm 1} \end{array}
\right],
\end{equation}

\begin{equation}
R_z\left(\frac{\pi }{{\rm 2}}\right)\left[ \begin{array}{c}
{\rm 1} \\ 
{\rm -}{\rm 1} \\ 
{\rm -}{\rm 1} \end{array}
\right]{\rm =}\left[ \begin{array}{ccc}
{\cos  \frac{\pi }{{\rm 2}}\ } & {\rm -}{\sin  \frac{\pi }{{\rm 2}}\ } & 0 \\ 
{\sin  \frac{\pi }{{\rm 2}}\ } & {\cos  \frac{\pi }{{\rm 2}}\ } & 0 \\ 
0 & 0 & {\rm 1} \end{array}
\right]\left[ \begin{array}{c}
{\rm 1} \\ 
{\rm -}{\rm 1} \\ 
{\rm -}{\rm 1} \end{array}
\right]{\rm =}\left[ \begin{array}{ccc}
0 & 0 & {\rm 1} \\ 
0 & {\rm 1} & 0 \\ 
{\rm -}{\rm 1} & 0 & 0 \end{array}
\right]\left[ \begin{array}{c}
{\rm 1} \\ 
{\rm -}{\rm 1} \\ 
{\rm -}{\rm 1} \end{array}
\right]{\rm =}\left[ \begin{array}{c}
{\rm 1} \\ 
{\rm 1} \\ 
{\rm -}{\rm 1} \end{array}
\right].
\end{equation}
\\
First it has been applied a rotation of $\frac{\pi }{{\rm 2}}$ about the x-axis which was followed by another $\frac{\pi }{{\rm 2}}$ rotation about the y-axis. At this point the 2 axis will be aligned and when rotating about the z-axis it is no longer possible to distinguish between ${\rm x}$ and ${\rm y}$. As result we lost one degree of freedom and this takes the name of $Gimbal$ $Lock$. According to the situation it is possible to opt for a different rotation order. This is an effective solution in some scenarios, for example when animating a camera in computer graphics: normally a camera will rotate only, or at least mostly, about two axis. By altering the rotation order is possible to cause the $Gimbal$ $Lock$ on the axis that is not being used but, with Euler angles, the $Gimbal$ $Lock$ will not disappear. 
\\
To further explore the previous example, below is a graphical and analytical representation of the different end-results with all the possible rotation orders. The order is usually written down as a sequence of the three axis, for example ${\rm zyx}$, where the hierarchy is represented bottom up starting from the left, so ${\rm x}$, in the ${\rm zyx}$ example, is the root parent. 
\begin{center}
\textit{\textcolor{red}{x = red}, \textcolor{green}{y = green}, \textcolor{blue}{z = blue}}\\

\begin{equation}
R_x\left(\frac{\pi }{2}\right)\left[ \begin{array}{c}
1 \\ 
1 \\ 
1 \end{array}
\right]\ \ \to \ \ \left[ \begin{array}{c}
1 \\ 
-1 \\ 
1 \end{array}
\right],
R_y\left(\frac{\pi }{2}\right)\ \ \to \ \left[ \begin{array}{c}
1 \\ 
-1 \\ 
-1 \end{array}
\right], 
R_z\left(\frac{\pi }{2}\right)\ \ \to \ \left[ \begin{array}{c}
1 \\ 
1 \\ 
-1 \end{array}
\right]
\end{equation}

\includegraphics[width=87.9mm, height=21.8mm]{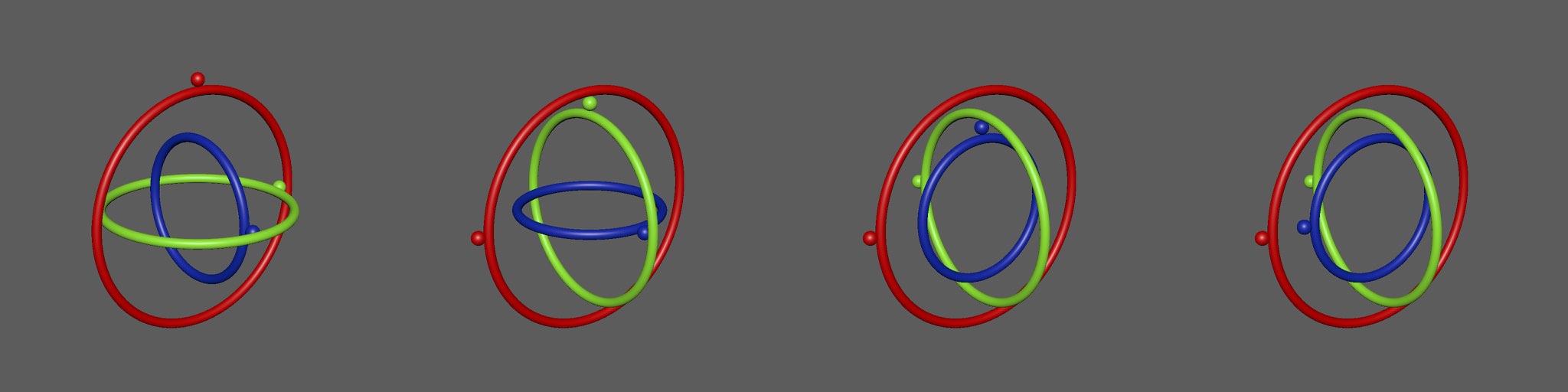}

Figure 1 - $Gimbal$ $Lock$, order: z y x

\begin{equation}
R_x\left(\frac{\pi }{2}\right)\left[ \begin{array}{c}
1 \\ 
1 \\ 
1 \end{array}
\right]\ \ \to \ \ \left[ \begin{array}{c}
1 \\ 
-1 \\ 
1 \end{array}
\right],
R_z\left(\frac{\pi }{2}\right)\ \ \to \ \ \left[ \begin{array}{c}
1 \\ 
1 \\ 
1 \end{array}
\right],
R_y\left(\frac{\pi }{2}\right)\ \ \to \left[ \begin{array}{c}
1 \\ 
1 \\ 
-1 \end{array}
\right]
\end{equation}

\includegraphics[width=87.9mm, height=21.8mm]{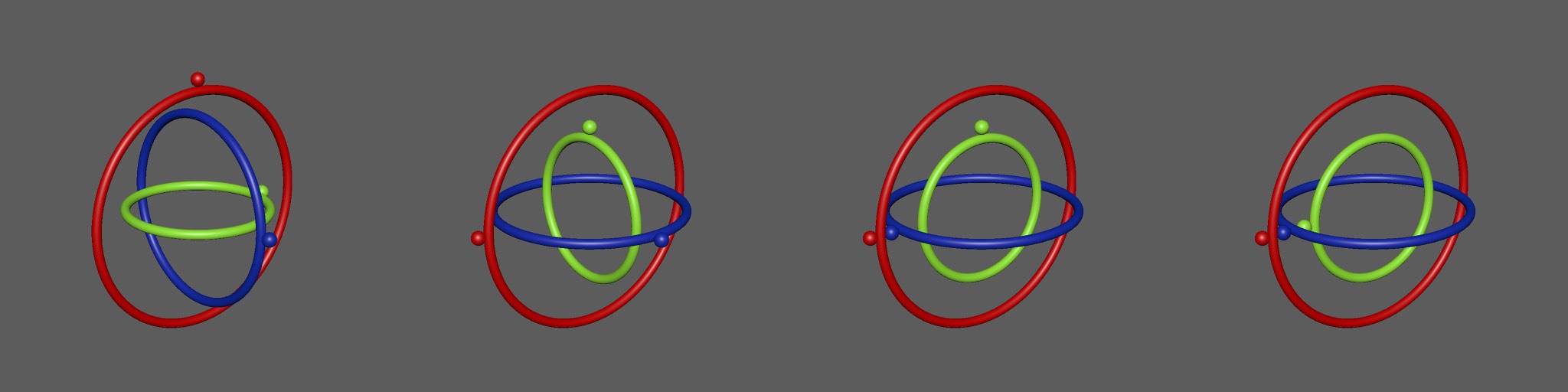}

Figure 2 - $Gimbal$ $Lock$, order: y z x.

\begin{equation}
R_y\left(\frac{\pi }{2}\right)\left[ \begin{array}{c}
1 \\ 
1 \\ 
1 \end{array}
\right]\ \ \to \ \ \left[ \begin{array}{c}
1 \\ 
1 \\ 
-1 \end{array}
\right],
R_x\left(\frac{\pi }{2}\right)\ \ \to \left[ \begin{array}{c}
1 \\ 
1 \\ 
1 \end{array}
\right],
R_z\left(\frac{\pi }{2}\right)\ \ \to \left[ \begin{array}{c}
-1 \\ 
1 \\ 
1 \end{array}
\right]
\end{equation}

\includegraphics[width=86.8mm, height=21.8mm]{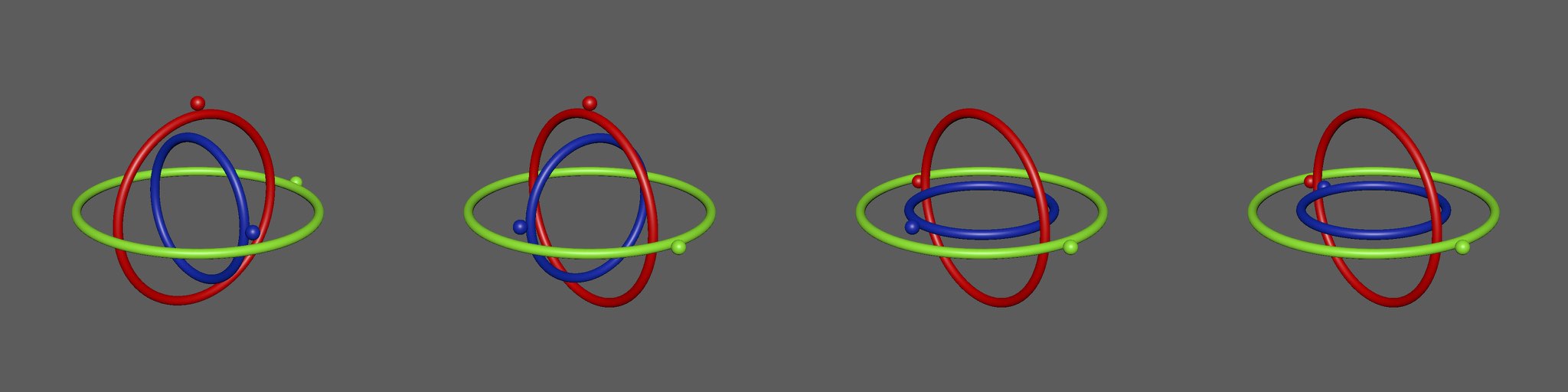}

Figure 3 - $Gimbal$ $Lock$, order: z x y.

\begin{equation}
R_y\left(\frac{\pi }{2}\right)\left[ \begin{array}{c}
1 \\ 
1 \\ 
1 \end{array}
\right]\ \ \to \ \ \left[ \begin{array}{c}
1 \\ 
1 \\ 
-1 \end{array}
\right],
R_z\left(\frac{\pi }{2}\right)\ \ \to \ \ \left[ \begin{array}{c}
-1 \\ 
1 \\ 
-1 \end{array}
\right],
R_x\left(\frac{\pi }{2}\right)\ \ \to \ \left[ \begin{array}{c}
-1 \\ 
1 \\ 
1 \end{array}
\right]
\end{equation}

\includegraphics[width=86.8mm, height=21.8mm]{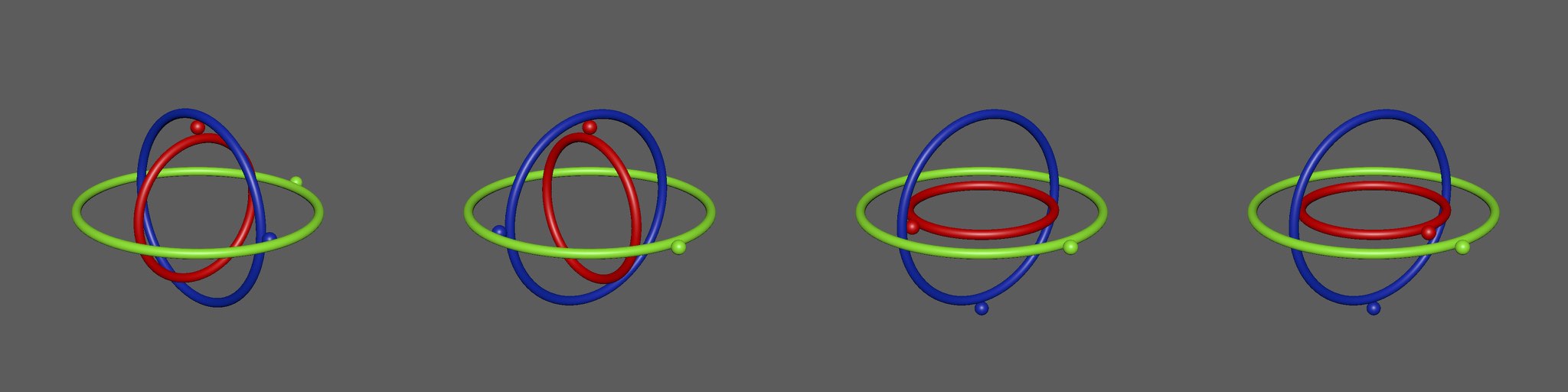}

Figure 4 - $Gimbal$ $Lock$, order: x z y.

\begin{equation}
R_z\left(\frac{\pi }{2}\right)\left[ \begin{array}{c}
1 \\ 
1 \\ 
1 \end{array}
\right]\ \ \to \ \left[ \begin{array}{c}
-1 \\ 
1 \\ 
1 \end{array}
\right],
R_x\left(\frac{\pi }{2}\right)\ \ \to \ \left[ \begin{array}{c}
-1 \\ 
-1 \\ 
1 \end{array}
\right],
R_y\left(\frac{\pi }{2}\right)\ \ \to \ \ \left[ \begin{array}{c}
1 \\ 
-1 \\ 
1 \end{array}
\right]
\end{equation}

\includegraphics[width=86.8mm, height=21.8mm]{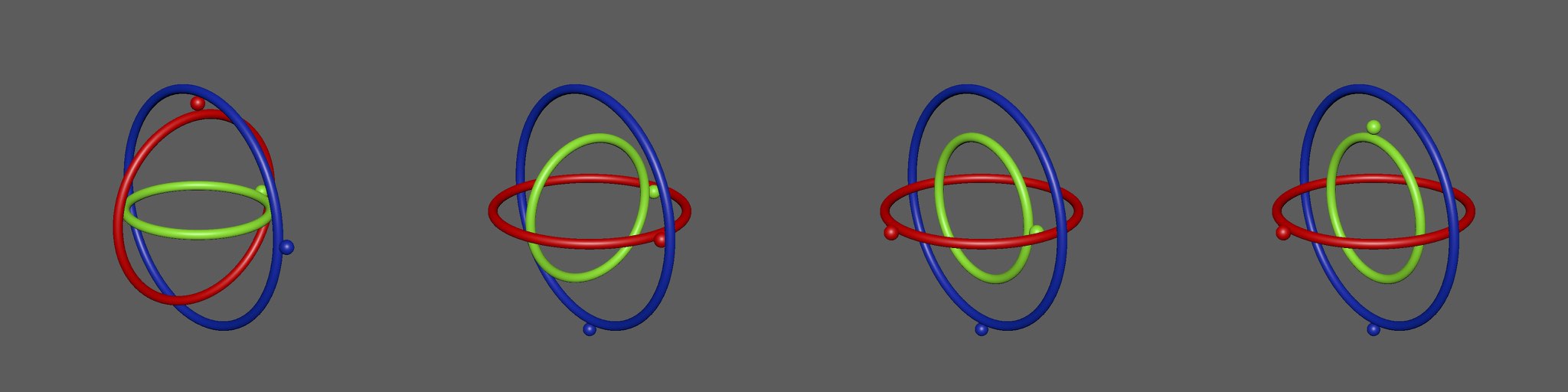}

Figure 5 - $Gimbal$ $Lock$, order: y x z.
\begin{equation}
R_z\left(\frac{\pi }{2}\right)\left[ \begin{array}{c}
1 \\ 
1 \\ 
1 \end{array}
\right]\ \ \to \ \left[ \begin{array}{c}
-1 \\ 
1 \\ 
1 \end{array}
\right],
R_y\left(\frac{\pi }{2}\right)\ \ \to \ \left[ \begin{array}{c}
1 \\ 
1 \\ 
1 \end{array}
\right],
R_x\left(\frac{\pi }{2}\right)\ \ \to \ \left[ \begin{array}{c}
1 \\ 
-1 \\ 
1 \end{array}
\right]
\end{equation}

\includegraphics[width=86.8mm, height=21.8mm]{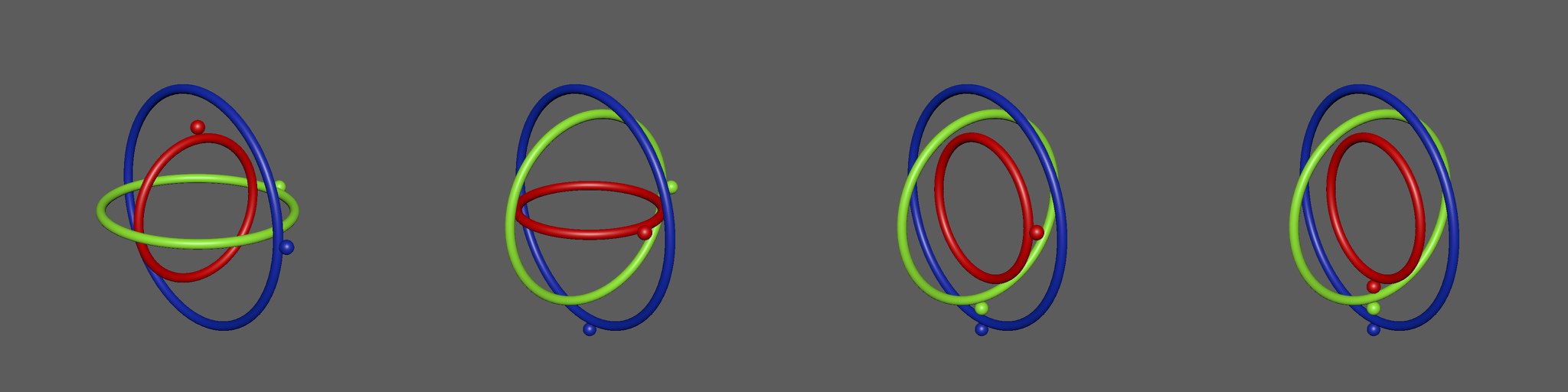}

Figure 6 - $Gimbal$ $Lock$, order: x y z.

\end{center}
The representation given above is basically a stylized gyroscope, which is a tool composed by a set of three gimbals of increasingly smaller diameter each of which is attached onto the above gimbal by two pivots that let it free to rotate on one axis only. The gyroscope is used in aerospace crafting to detect the three main motions of an aircraft: pitch, roll and yaw.  
\\ 
Due to its structure, also the gyroscope can encounter the $Gimbal$ $Lock$ with catastrophic consequences. The most famous example of a real life $Gimbal$ $Lock$ situation is definitely the Apollo 10 accident \cite{Han}. 

\subsection{\textit{Topological Reasons}}

The group of all the rotations about the origin of a three-dimensional Euclidean space ${{\mathbb R}}^{{\rm 3}}$ is usually denoted as $SO(3)$. Since the rotations are linear transformation of ${{\mathbb R}}^{{\rm 3}}$ and can be represented with matrix multiplications, it is possible to use orthonormal basis of  ${{\mathbb R}}^{{\rm 3}}$. Such matrices are called $special$ $orthonormal$ $matrices$, which motivates the name $SO(3)$. 
\\
To understand why the $Gimbal$ $Lock$ happens the first step is to visualize the ball of three dimensions in ${{\mathbb R}}^{{\rm 3}}$, this ball will be the set of point having distance $\rho $ from the origin. Each of these points, if connected to the origin, creates an axis. The rotation of $-\pi $ and $\pi {\rm \ }$actually individuate the same point but on the opposite side of the origin, which therefore is a glued-together antipodal point. 
\\
The ball just constructed is diffeomorphic to the real projective space ${RP}^{{\rm 3}}$, which, by definition, is the topological space of lines passing through the origin 0 in ${{\mathbb R}}^{n{\rm +1}}$. 
\noindent
\\
If the ray $\rho $ of the ball is 1, the ball in ${{\mathbb R}}^{{\rm 3}}$ becomes a unit 3-sphere denoted as $S^{{\rm 3}}$. Since both $S^{{\rm 3}}$ and $SO(3)$ are diffeomorphic to ${RP}^{{\rm 3}}$, then $S^{{\rm 3}}$ is diffeomorphic to $SO(3)$.
\noindent
\\
Euler rotations consist in circles multiplications but, since circle multiplications, in topology, are equal to a torus, it is possible to say that:
\begin{equation}
S^{{\rm 1}}{\rm \times }S^{{\rm 1}}{\rm \times }S^{{\rm 1}}\ne S^{{\rm 3}}
\end{equation}
\noindent
However Euler rotations can be mapped on to $S^{{\rm 3}}$ by an operation of suspension, by squashing, squeezing and pinching the torus but, unfortunately, this topology change is not harm free. 
\\
\\
The way to map ${\left(S^{{\rm 1}}\right)}^{{\rm 3}}$ onto $S^{{\rm 3}}$ is to squash down one of the 3 circles until it gets a linear interval. The singularities also known as $Gimbal$ $Lock$ are the ends of this linear intervals each of which, even though looks like $S^{{\rm 1}}{\rm \times }S^{{\rm 1}}$, it actually only is $S^{{\rm 1}}$ and need to be collapsed down 1 dimension. 
\\
\\
Since $SO(3)$ is diffeomorphic to $S^{{\rm 3}}$ and ${\left(S^{{\rm 1}}\right)}^{{\rm 3}}\ne S^{{\rm 3}}$ then ${\left(S^{{\rm 1}}\right)}^{{\rm 3}}$ is topologically different from $SO(3)$. This proves intuitively that Euler angles can't be mapped nicely in a one-to-one way onto the rotation group. 
\\
\\
The group $S^{3}$ was constructed starting from the ball of ray 1. From an algebraic point of view this can be interpreted as the group of Quaternions whose absolute value is 1. Quaternions are a key concept in understanding rotations.

\section{Quaternions and rotations}
\subsection{\textit{Exponential Map}}

Since this property is somehow inherited from complex numbers, it is easy to introduce this starting from there. Johnson \cite{Joh} starts by applying the definition of logarithm to the following expression:

\begin{equation}
{\log  \left(e^{i\theta }\right)\ }{\rm =}i\theta,
\end{equation}
\\
which is true up to a factor of ${\rm 2}k\pi $:
\begin{equation}
e^{i{\rm (}\theta {\rm +2}k\pi {\rm )}}{\rm =}e^{i\theta },
\end{equation}
\\
with $k$ being an integer. He then proceeds by investigating the infinitesimal transformations of the complex number and their relation to the logarithm. After expressing the exponential as a power series and applying DeMoivre it is possible to compute the derivatives, which leads to:
\begin{equation}
\frac{d{\log  \left(e^{it\theta }\right)\ }}{dt}{\rm =}i\theta
\end{equation}
\\
It is obvious from the results just found that the $\textit{log}$ function creates a linear space (since the derivative is constant in $dt$) which acts as an invertible map over the complex numbers. 
\\
More precisely, the logarithm can be thought as non-periodic pure imaginary number representable on a real line.
\\
On the other end, the exponential is curved and periodic as a result of the Euler's formula. Furthermore, there is a correspondence between the flat space of the logarithm and the exponential, given by the mapping of each point of the logarithm to the circle $S^{{\rm 1}}$. 

\begin{center}
\includegraphics[width=50.3mm, height=40.0mm]{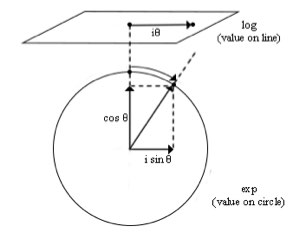}

Figure \textit{7} - Correspondance between logarithm flat space and exponential curved space.
\end{center}
From (3.2) we can deduce that the same point on the circle can be mapped to an infinite amount of points in the logarithm. 
\\
\\
\textbf{Definition.} \textit{The exponential map of a pure imaginary number produces a circle. The logarithm of a point on the circle is not unique but many-valued.}
\\
\\
This mathematical tool becomes much more powerful in the case of quaternions as it allows to map 3D-vectors into Quaternions and viceversa. The “viceversa” implies that with the log map is possible to locally linearize quaternions in an invertible way. This is intimately connected to the concepts of tangent space and Lie Algebra \cite{Sa We}.
\\
One possible proof is analogous to the one for the complex numbers and it is here omitted. The main difference is that in the case of the complex numbers, the tangent space was a 1D flat space. With Quaternions, the tangent space is a hyperplane of three dimensions. Let’s try to visualize it as it follows:

\begin{center}
\includegraphics[width=89.0mm, height=42.4mm]{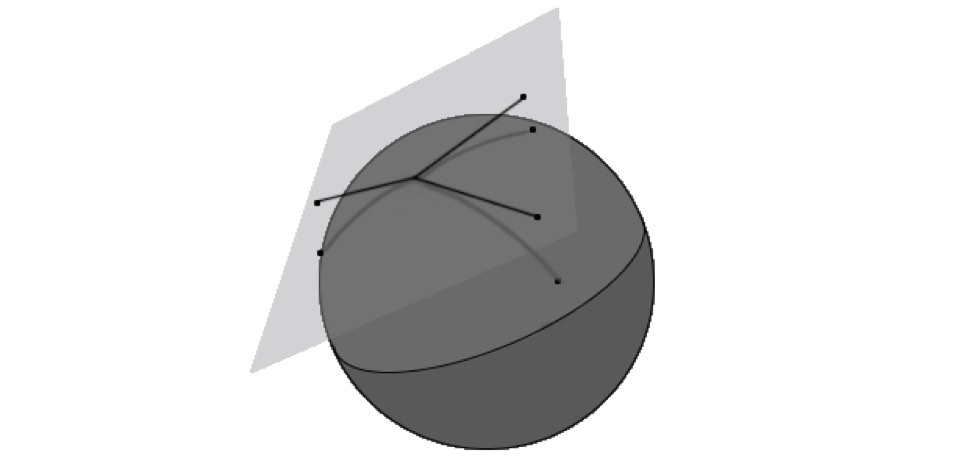}

Figure \textit{8} - Quaternions exponential map and tangent space.
\end{center}

\subsection{\textit{Interpolation and Gimbal Lock}}
It is well known that Quaternions have different ways of being visualized, whether it is from a geometrical or algebraic point of view, or from an interpolation or calculus perspective. The interpolation of Quaternions is an extension of the interpolation of polynomials in the Euclidean space.  Let’s start by giving a definition of interpolation and try to extend it up to the Quaternions following Hanson \cite{Han}:
\\
\\
\textbf{Definition} \textbf{(Interpolation).} \textit{Interpolation is an estimation of a value within two known values.}
\\
\\
For example, if we have two points on a line, i.e. $x_0{\rm =}x\left(0\right)$ and $x_{{\rm 1}}{\rm =}x\left({\rm 1}\right)$, we can interpolate the values in between using a parameter here denoted $t$.

\begin{center}
\includegraphics[width=85.1mm, height=18.9mm]{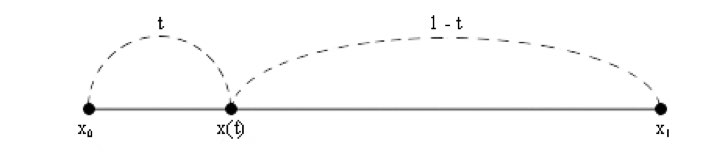}

Figure \textit{9} - Lerp.
\end{center}
\noindent
We can write this operation, known as Linear Interpolation and often abbreviated with $Lerp$, as follow:
\begin{equation}
Lerp{\rm (}x_0{\rm ,\ }x_{{\rm 1}};;t{\rm )=}x_0{\rm +}t{\rm (}x_{{\rm 1}}{\rm -}x_0{\rm )=}\left({\rm 1-}t\right)x_0{\rm +}tx_{{\rm 1}}
\end{equation}
\noindent
In order to restrict the parametric curve to be between $x_0$ and $x_{{\rm 1}}$, the parameter $t$ must be ${\rm 0}\le t\le {\rm 1}$.
\\
Unfortunately this doesn't make much sense when extended to a sphere. Given two points on the sphere, namely $q_0$ and $q_{{\rm 1}}$, even though their norm is the same, if interpolated with $Lerp$ the in-betweens can assume any length as shown in the picture below:

\begin{center}
\includegraphics[width=88.8mm, height=36.3mm]{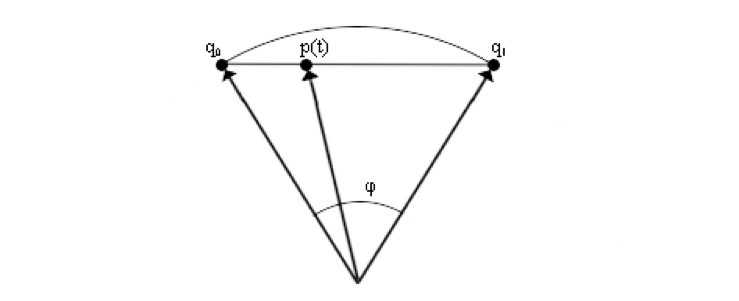}

Figure \textit{10} - Lerp on sphere.
\end{center}

\begin{center}
\includegraphics[width=88.8mm, height=36.3mm]{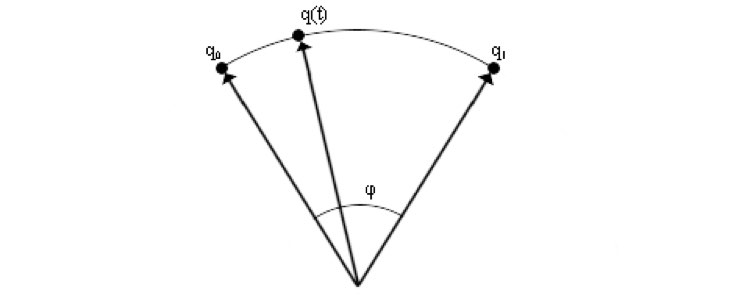}

Figure \textit{11} - Expected interpolation on the sphere,
\end{center}
where:

\begin{equation}
p\left(t\right){\rm =}\left({\rm 1-}t\right)q_0{\rm +}tq_{{\rm 1}}
\end{equation}
\\
Let's now write the following expression and compute by keeping in mind that $q_0\cdot q_{{\rm 1}}{\rm =}{\cos  \varphi \ }$:
\begin{equation}
p\left(t\right)\cdot p\left(t\right){\rm =1-2}t{\rm +2}t^{{\rm 2}}{\rm +2}t{\rm (1-}t{\rm )}{\cos  \varphi \ }
\end{equation}
\\
Obviously the term ${\cos  \varphi \ }$ just introduced proves that the length is not fixed and the point $p\left(t\right)$ will not stay on the circle as we want.
\\
Even though is possible to normalize the norm of the point interpolated with $Lerp$, that wouldn't satisfy the condition of having constant velocity in the interpolation. This means that in the sphere we are looking for a constant angular velocity. 
\\
We are now interested in deriving a new formulation of $Lerp$ that adapts to spheres automatically by adjusting the norm and providing a constant angular velocity. This new method is called $Slerp$ which stands for Spherical Linear Interpolation.
\\
The operation of Spherical Linear Interpolation (following the Gram-Schmidt \cite{Han} derivation) between two points $q_0$ and $q_{{\rm 1}}$ belonging to a sphere (circle) according to a parameter $t$ such that ${\rm 0}\le t\le {\rm 1}$, is defined as:
\\
\begin{equation}
Slerp{\rm (}q_0{\rm ,\ }q_{{\rm 1}};;t{\rm )=}q_0\frac{{\sin  {\rm (1-}t{\rm )}\varphi \ }}{{\sin  \varphi \ }}{\rm +}q_{{\rm 1}}\frac{{\sin  t\varphi \ }}{{\sin  \varphi \ }}
\end{equation}

\begin{center}
\includegraphics[width=88.7mm, height=60.5mm]{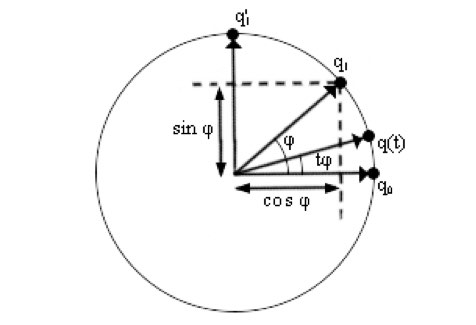}

Figure 12 - Slerp.
\end{center}
\noindent
The interesting property of $Slerp$ is that it carries on any dimension since it only rotates a vector on the plane spanned by the start and end position. For this reason, let's see how to extend it to Quaternions. 
\\
In order to have the $Slerp$ to work on Quaternions, we need to remap some of the linear interpolation operations to Quaternions interpolation operations as explained by Johnson \cite{Joh}.
\\
Therefore  the operation of Spherical Linear Interpolation between two unit Quaternions, according to a parameter $t$ such that ${\rm 0}\le t\le {\rm 1}$, is defined as:
\begin{equation}
Slerp\left(q_0{\rm ,\ }q_{{\rm 1}};t\right){\rm =}q_0{\left(q^{{\rm -}{\rm 1}}_0q_{{\rm 1}}\right)}^t
\end{equation}
\\
At this point we have all the tools to discuss how Quaternions can prevent $Gimbal$ $Lock$. The first question we need to answer is: is it really true Quaternions do not suffer of $Gimbal$ $Lock$? The answer is no, that is not precise. 
\\
Let's imagine to apply a transformation defined as a sequence of rotations to an object in such a way to cause a $Gimbal$ $Lock$:
\begin{equation}
T\left(\theta {\rm ,\ }\varphi {\rm ,\ }\psi \right){\rm =}R{\rm (}\theta {\rm ,\ }\hat{x}{\rm )}\circ R{\rm (}\varphi {\rm ,\ }\hat{y}{\rm )}\circ R{\rm (}\psi {\rm ,\ }\hat{z}{\rm )}
\end{equation}
\\
If we translate this 1-to-1 to Quaternions, we get:
\begin{equation}
Q\left(\theta {\rm ,\ }\varphi {\rm ,\ }\psi \right){\rm =}e^{\hat{x}\theta }e^{\hat{y}\varphi }e^{\hat{z}\psi }
\end{equation}
\\
It is easy to verify by plotting that if $T$ was in $Gimbal$ $Lock$, $Q$ will also  result in a $Gimbal$ $Lock$. 
\\
The way Quaternions can rescue us is by representing this sequence of rotations as a whole using Euler's rotation theorem. The problem is that it is not intuitive to find the right axis to parametrize the rotation but it can be recovered from the rotational matrix. 
\\
To do so, we know that the rotation around the basis $\hat{x}$, $\hat{y}$ and $\hat{z}$ can be represented by the matrices:
\begin{align}
R_x\left(\varphi \right){\rm =}\left[ \begin{array}{ccc}
{\rm 1} & 0 & 0 \\ 
0 & cos{\rm \ }\varphi  & {\rm -}sin\varphi  \\ 
0 & sin\varphi  & cos{\rm \ }\varphi  \end{array}
\right]\\
R_y\left(\varphi \right){\rm =}\left[ \begin{array}{ccc}
cos{\rm \ }\varphi  & 0 & sin\varphi  \\ 
0 & {\rm 1} & 0 \\ 
{\rm -}sin\varphi  & 0 & cos{\rm \ }\varphi  \end{array}
\right]\\
R_z\left(\varphi \right){\rm =}\left[ \begin{array}{ccc}
cos{\rm \ }\varphi  & {\rm -}sin\varphi  & 0 \\ 
sin\varphi  & cos{\rm \ }\varphi  & 0 \\ 
0 & 0 & {\rm 1} \end{array}
\right] 
\end{align}
\\
In order to find the equivalent axis-angle representation $R{\rm (}\theta {\rm ,\ }\hat{n}{\rm )}$, we can define $\hat{n}$ in terms of polar coordinates:
\\
\begin{equation}
\hat{n}{\rm =}\left(cos{\rm \ }\alpha {\rm \ }sin{\rm \ }\beta {\rm ,\ }sin{\rm \ }\alpha {\rm \ }sin{\rm \ }\beta {\rm ,\ }cos{\rm \ }\beta \right), with {\rm 0}\le \alpha \le {\rm 2}\pi , {\rm 0}\le \beta \le \pi
\end{equation}
\\
The rest of the process is very verbose and here is omitted but can be found on \cite{Joh, Pal B R, Dol}. This process will yield to:
\begin{equation}
R\left(\theta {\rm ,\ }\hat{n}\right){\rm =}\left[ \begin{array}{ccc}
c{\rm +}{n_{{\rm 1}}}^{{\rm 2}}{\rm (1-}c{\rm )} & n_{{\rm 1}}n_{{\rm 2}}{\rm (1-}c{\rm )-}sn_{{\rm 3}} & n_{{\rm 1}}n_{{\rm 3}}{\rm (1-}c{\rm )+}sn_{{\rm 2}} \\ 
n_{{\rm 2}}n_{{\rm 1}}{\rm (1-}c{\rm )+}sn_{{\rm 3}} & c{\rm +}{n_{{\rm 2}}}^{{\rm 2}}{\rm (1-}c{\rm )} & n_{{\rm 2}}n_{{\rm 3}}{\rm (1-}c{\rm )-}sn_{{\rm 1}} \\ 
n_{{\rm 3}}n_{{\rm 1}}{\rm (1-}c{\rm )-}sn_{{\rm 2}} & n_{{\rm 3}}n_{{\rm 2}}{\rm (1-}c{\rm )+}sn_{{\rm 1}} & c{\rm +}{n_{{\rm 3}}}^{{\rm 2}}{\rm (1-}c{\rm )} \end{array}
\right] 
\end{equation}
\\
From this matrix just found, that we will call $R$ for simplicity, we can derive both the axis and the angle as follow.
\\
In order to derive the axis let's find the skew-symmetric matrix:

\begin{equation}
R{\rm -}R^T{\rm =}\left[ \begin{array}{ccc}
0 & {\rm -}{\rm 2}sn_{{\rm 3}} & {\rm 2}sn_{{\rm 2}} \\ 
{\rm 2}sn_{{\rm 3}} & 0 & {\rm -}{\rm 2}sn_{{\rm 1}} \\ 
{\rm -}{\rm 2}sn_{{\rm 2}} & {\rm 2}sn_{{\rm 1}} & 0 \end{array}
\right]{\rm =}\left[ \begin{array}{ccc}
0 & {\rm -}c & b \\ 
c & 0 & {\rm -}a \\ 
{\rm -}b & a & 0 \end{array}
\right]
\end{equation}
\\
We can then compute $d$ as:
\begin{equation}
d{\rm =}\sqrt{a^{{\rm 2}}{\rm +}b^{{\rm 2}}{\rm +}c^{{\rm 2}}} 
\end{equation}
\\
It follows after normalization that:
\begin{equation}
\hat{n}{\rm =(}\frac{a}{d}{\rm ,\ }\frac{b}{d}{\rm ,\ }\frac{c}{d}{\rm )}
\end{equation}
\\
As for the angle, we know that:

\[Tr\left(R\right){\rm =\ }c{\rm +}{n_{{\rm 1}}}^{{\rm 2}}\left({\rm 1-}c\right){\rm +\ }c{\rm +}{n_{{\rm 2}}}^{{\rm 2}}\left({\rm 1-}c\right){\rm +}c{\rm +}{n_{{\rm 3}}}^{{\rm 2}}\left({\rm 1-}c\right)\] 
\[{\rm =3}c{\rm +}\left({\rm 1-}c\right)\left({n_{{\rm 1}}}^{{\rm 2}}{\rm +}{n_{{\rm 2}}}^{{\rm 2}}{\rm +}{n_{{\rm 3}}}^{{\rm 2}}\right)\]
\begin{equation}
{=1+2}c{\rm =1+2}{\cos  \theta \ }
\end{equation}
\\
Therefore:
\begin{equation}
{\cos \theta }{\rm =}\frac{Tr\left(R\right){\rm -}{\rm 1}}{{\rm 2}}
\end{equation}

\begin{equation}
{\sin\theta  }{\rm =}\sqrt{{\rm 1-}{\left(\frac{Tr\left(R\right){\rm -}{\rm 1}}{{\rm 2}}\right)}^{{\rm 2}}}
\end{equation}
\\
\textbf{Remark}. This process is not always valid, for example if $\theta = \pi $ the skew symmetric matrix is not going to be helpful. For a full overview of the different cases and possible solutions, the interested reader  refers to \cite{Sh Gr M N}.
\\
We conclude this paragraph with an observation. Eventually we will be interested in how to work and compute Quaternions from a programming point of view. Computationally wise one might be misled to think Quaternions are faster than matrices since a 3x3 matrix (often 4x4 in computer graphics packages) has to store and work with 9 (or 16) values rather than 4 as in the Quaternions case. This is not entirely true because of the time required to convert from rotation matrices to Quaternions or the other way around, so it really depends on the specific case whether computation will be faster with Quaternions or not. For a full analysis it is recommended to read the article by Eberly \cite{Ebe}.
\section{Creation of a RBF solver for linear blending}

\subsection{\textit{The mathematical machinery}}

Let's start by calling the $K{\rm \times }N$ input matrix $T$, and the $K{\rm \times }M$ samples matrix $S$. The goal is, given a current position, to interpolate the samples. In order to do so, we will need a matrix of weights, named $W$, that will depend on the distance between the current position, which will be a ${\rm 1\times }N$ matrix that we will call $L$, and each row (position) in $T$.  
\\
If we call${\rm \ }\xi $ the matrix of the distances between each row of $T$ and $L$, and $\Omega $ the ${\rm 1\times }M$ output matrix (after interpolation), the problem can be written as:

\begin{equation}
\xi W{\rm =}\Omega 
\end{equation}
\\
Where $W$ can be found as follow:

\[DW{\rm =}S\]
\[D^{{\rm -}{\rm 1}}DW{\rm =}D^{{\rm -}{\rm 1}}S\]
\begin{equation}
W{\rm =}D^{{\rm -}{\rm 1}}S
\end{equation}
\\
With $D$ being the symmetric${\rm \ }K{\rm \times }K$ distance matrix of the rows of $T$ taken pairwise. It follows: 

\begin{equation}
\xi D^{{\rm -}{\rm 1}}S{\rm =}\Omega
\end{equation}
\\
So far the radial basis functions were not mentioned. So, what do we use those functions for? They are used to calculate the distances. Before going any further let's have a look at some of the main radial basis functions \cite{Buh, Roh}: 
\\
\\
Gaussian: 
\begin{equation}
\varphi {\rm (}r{\rm )=}e^{{\rm -}{{\rm (}\varepsilon r{\rm )}}^{{\rm 2}}} 
\end{equation}
\\
Multiquadratic: 
\begin{equation}
\varphi {\rm (}r{\rm )=}\sqrt{{\rm 1+}{{\rm (}\varepsilon r{\rm )}}^{{\rm 2}}}
\end{equation}
\\
Inverse quadratic: 
\begin{equation}
\varphi {\rm (}r{\rm )=}\frac{{\rm 1}}{{\rm 1+}{{\rm (}\varepsilon r{\rm )}}^{{\rm 2}}}
\end{equation}
\\
Inverse multiquadratic: 
\begin{equation}
\varphi {\rm (}r{\rm )=}\frac{{\rm 1}}{\sqrt{{\rm 1+}{{\rm (}\varepsilon r{\rm )}}^{{\rm 2}}}}
\end{equation}
\\
Polyharmonic:

\begin{equation}
\left\{ \begin{array}{c}
\varphi \left(r\right) = r^{\varepsilon }\,\,\,\,\,  if \,\varepsilon \in \{ 2n +1, n\in {\mathbb N} \} \\ 
\varphi \left(r\right) = r^{\varepsilon }{\ln  (r)\ }\,\,\,     if\, \varepsilon \in \left\{2n. n\in {\mathbb N}\right\}
\end{array}
\right.
\end{equation}
\\
Thinplate:
\\
\begin{equation}
\varphi \left(r\right){\rm =}r^{{\rm 2}}{\ln(r)\ }
\end{equation}
\\
Those are the main functions that can be used to perform function approximation. Such approximation is written as the sum of $K$ radial basis function calculated from different centers $r_i$ and weighted by $w_i$:

\begin{equation}
f\left(r\right){\rm =}\sum^K_{i{\rm =1}}{w_i\varphi {\rm (}\left\|r{\rm -}r_i\right\|{\rm )}}
\end{equation}
\\
Now that the general idea has been presented, let's go a little bit more into details and start from defining $T$ and $S$:

\begin{align}
T{\rm =}\left[ \begin{array}{ccc}
 \begin{array}{c}
t_{{\rm 1,1}} \\ 
 \begin{array}{c}
t_{{\rm 2,1}} \\ 
\vdots  \\ 
t_{k{\rm -}{\rm 1,1}} \end{array}
 \\ 
t_{k{\rm ,1}} \end{array}
 &  \begin{array}{ccc}
 \begin{array}{c}
t_{{\rm 1,2}} \\ 
 \begin{array}{c}
t_{{\rm 2,2}} \\ 
\vdots  \\ 
t_{k{\rm -}{\rm 1,2}} \end{array}
 \\ 
t_{k{\rm ,2}} \end{array}
 &  \begin{array}{c}
{\rm \dots } \\ 
 \begin{array}{c}
 \\ 
\ddots  \\ 
 \end{array}
 \\ 
{\rm \dots } \end{array}
 &  \begin{array}{c}
t_{{\rm 1,}N{\rm -}{\rm 1}} \\ 
 \begin{array}{c}
t_{{\rm 2,}N{\rm -}{\rm 1}} \\ 
\vdots  \\ 
t_{k{\rm -}{\rm 1,}N{\rm -}{\rm 1}} \end{array}
 \\ 
t_{k,N{\rm -}{\rm 1}} \end{array}
 \end{array}
 &  \begin{array}{c}
t_{{\rm 1,}N} \\ 
 \begin{array}{c}
t_{{\rm 2,}N} \\ 
\vdots  \\ 
t_{k{\rm -}{\rm 1,}N} \end{array}
 \\ 
t_{k,N} \end{array}
 \end{array}
\right]
\end{align}

\begin{align}
S{\rm =}\left[ \begin{array}{ccc}
 \begin{array}{c}
s_{{\rm 1,1}} \\ 
 \begin{array}{c}
s_{{\rm 2,1}} \\ 
\vdots  \\ 
s_{k{\rm -}{\rm 1,1}} \end{array}
 \\ 
s_{k{\rm ,1}} \end{array}
 &  \begin{array}{ccc}
 \begin{array}{c}
s_{{\rm 1,2}} \\ 
 \begin{array}{c}
s_{{\rm 2,2}} \\ 
\vdots  \\ 
s_{k{\rm -}{\rm 1,2}} \end{array}
 \\ 
s_{k{\rm ,2}} \end{array}
 &  \begin{array}{c}
{\rm \dots } \\ 
 \begin{array}{c}
 \\ 
\ddots  \\ 
 \end{array}
 \\ 
{\rm \dots } \end{array}
 &  \begin{array}{c}
s_{{\rm 1,}M{\rm -}{\rm 1}} \\ 
 \begin{array}{c}
s_{{\rm 2,}M{\rm -}{\rm 1}} \\ 
\vdots  \\ 
s_{k{\rm -}{\rm 1,}M{\rm -}{\rm 1}} \end{array}
 \\ 
s_{k,M{\rm -}{\rm 1}} \end{array}
 \end{array}
 &  \begin{array}{c}
s_{{\rm 1,}M} \\ 
 \begin{array}{c}
s_{{\rm 2,}M} \\ 
\vdots  \\ 
s_{k{\rm -}{\rm 1,}M} \end{array}
 \\ 
s_{k,M} \end{array}
 \end{array}
\right]
\end{align}
\\
\\
Next, let's define the distance matrix $D$ as:

\setlength{\parskip}{18.0pt}
 \[ D= \left[ \begin{matrix}
\begin{matrix}
 \varphi  \left(  \Vert t_{1,\ast}-t_{1,\ast} \Vert  \right) \\
\begin{matrix}
 \varphi  \left(  \Vert t_{2,\ast}-t_{1,\ast} \Vert  \right) \\
 \vdots \\
 \varphi  \left(  \Vert t_{K-1,\ast}-t_{1,\ast} \Vert  \right) \\
\end{matrix}
\\
 \varphi  \left(  \Vert t_{K,\ast}-t_{1,\ast} \Vert  \right) \\
\end{matrix}
  &  \begin{matrix}
\begin{matrix}
 \varphi  \left(  \Vert t_{1,\ast}-t_{2,\ast} \Vert  \right) \\
\begin{matrix}
 \varphi  \left(  \Vert t_{2,\ast}-t_{2,\ast} \Vert  \right) \\
 \vdots \\
 \varphi  \left(  \Vert t_{K-1,\ast}-t_{2,\ast} \Vert  \right) \\
\end{matrix}
\\
 \varphi  \left(  \Vert t_{K,\ast}-t_{2,\ast} \Vert  \right) \\
\end{matrix}
  &  \begin{matrix}
 \ldots \\
\begin{matrix}
\\
\ddots\\
\\
\end{matrix}
\\
 \ldots \\
\end{matrix}
  &  \begin{matrix}
 \varphi  \left(  \Vert t_{1,\ast}-t_{K-1,\ast} \Vert  \right) \\
\begin{matrix}
 \varphi  \left(  \Vert t_{2,\ast}-t_{K-1,\ast} \Vert  \right) \\
 \vdots \\
 \varphi  \left(  \Vert t_{K-1,\ast}-t_{K-1,\ast} \Vert  \right) \\
\end{matrix}
\\
 \varphi  \left(  \Vert t_{K,\ast}-t_{K-1,\ast} \Vert  \right) \\
\end{matrix}
\\
\end{matrix}
  &  \begin{matrix}
 \varphi  \left(  \Vert t_{1,\ast}-t_{K,\ast} \Vert  \right) \\
\begin{matrix}
 \varphi  \left(  \Vert t_{2,\ast}-t_{K,\ast} \Vert  \right) \\
 \vdots \\
 \varphi  \left(  \Vert t_{K-1,\ast}-t_{K,\ast} \Vert  \right) \\
\end{matrix}
\\
 \varphi  \left(  \Vert t_{K,\ast}-t_{K,\ast} \Vert  \right) \\
\end{matrix}
\\
\end{matrix}
 \right] \] \par

\vspace{\baselineskip}
 \[ = \left[ \begin{matrix}
\begin{matrix}
0\\
\begin{matrix}
 \varphi  \left(  \Vert t_{2,\ast}-t_{1,\ast} \Vert  \right) \\
 \vdots \\
 \varphi  \left(  \Vert t_{K-1,\ast}-t_{1,\ast} \Vert  \right) \\
\end{matrix}
\\
 \varphi  \left(  \Vert t_{K,\ast}-t_{1,\ast} \Vert  \right) \\
\end{matrix}
  &  \begin{matrix}
\begin{matrix}
 \varphi  \left(  \Vert t_{1,\ast}-t_{2,\ast} \Vert  \right) \\
\begin{matrix}
0\\
 \vdots \\
 \varphi  \left(  \Vert t_{K-1,\ast}-t_{2,\ast} \Vert  \right) \\
\end{matrix}
\\
 \varphi  \left(  \Vert t_{K,\ast}-t_{2,\ast} \Vert  \right) \\
\end{matrix}
  &  \begin{matrix}
 \ldots \\
\begin{matrix}
\\
\ddots\\
\\
\end{matrix}
\\
 \ldots \\
\end{matrix}
  &  \begin{matrix}
 \varphi  \left(  \Vert t_{1,\ast}-t_{K-1,\ast} \Vert  \right) \\
\begin{matrix}
 \varphi  \left(  \Vert t_{2,\ast}-t_{K-1,\ast} \Vert  \right) \\
 \vdots \\
0\\
\end{matrix}
\\
 \varphi  \left(  \Vert t_{K,\ast}-t_{K-1,\ast} \Vert  \right) \\
\end{matrix}
\\
\end{matrix}
  &  \begin{matrix}
 \varphi  \left(  \Vert t_{1,\ast}-t_{K,\ast} \Vert  \right) \\
\begin{matrix}
 \varphi  \left(  \Vert t_{2,\ast}-t_{K,\ast} \Vert  \right) \\
 \vdots \\
 \varphi  \left(  \Vert t_{K-1,\ast}-t_{K,\ast} \Vert  \right) \\
\end{matrix}
\\
0\\
\end{matrix}
\\
\end{matrix}
 \right] \] \par

\begin{equation}
{\rm =}\left[ \begin{array}{ccc}
 \begin{array}{c}
0 \\ 
 \begin{array}{c}
d_{{\rm 2,1}} \\ 
\vdots  \\ 
d_{k{\rm -}{\rm 1,1}} \end{array}
 \\ 
d_{k{\rm ,1}} \end{array}
 &  \begin{array}{ccc}
 \begin{array}{c}
d_{{\rm 1,2}} \\ 
 \begin{array}{c}
0 \\ 
\vdots  \\ 
d_{k{\rm -}{\rm 1,2}} \end{array}
 \\ 
d_{k{\rm ,2}} \end{array}
 &  \begin{array}{c}
{\rm \dots } \\ 
 \begin{array}{c}
 \\ 
\ddots  \\ 
 \end{array}
 \\ 
{\rm \dots } \end{array}
 &  \begin{array}{c}
d_{{\rm 1,}K{\rm -}{\rm 1}} \\ 
 \begin{array}{c}
d_{{\rm 2,}K{\rm -}{\rm 1}} \\ 
\vdots  \\ 
0 \end{array}
 \\ 
d_{k,K{\rm -}{\rm 1}} \end{array}
 \end{array}
 &  \begin{array}{c}
d_{{\rm 1,}K} \\ 
 \begin{array}{c}
d_{{\rm 2,}K} \\ 
\vdots  \\ 
d_{k{\rm -}{\rm 1,}K} \end{array}
 \\ 
0 \end{array}
 \end{array}
\right]
\end{equation}
\\
\\
We immediately notice that $D$ is a $K{\rm \times }K$ square symmetric matrix  with all 0 on the main diagonal (because the distance of a row from itself is 0). 
\\
\\
\textbf{Example.} Let's see how the element  $d_{x,y}$, with ${\rm 0<}x{\rm <}K$ and ${\rm 0<}y{\rm <}K$, is calculated. 

\[t_{x{\rm ,*}}{\rm =[} \begin{array}{ccc}
 \begin{array}{cc}
{\rm \ }t_{x{\rm ,1}} & t_{x{\rm ,2}} \end{array}
 & {\rm \dots } &  \begin{array}{cc}
t_{x,N{\rm -}{\rm 1}} & t_{x,N} \end{array}
 \end{array}
{\rm ]}\]\\
\[t_{y{\rm ,*}}{\rm =[} \begin{array}{ccc}
 \begin{array}{cc}
{\rm \ }t_{y{\rm ,1}} & t_{y{\rm ,2}} \end{array}
 & {\rm \dots } &  \begin{array}{cc}
t_{y,N{\rm -}{\rm 1}} & t_{y,N} \end{array}
 \end{array}
{\rm ]}\]

\begin{equation}
\left\|t_{x{\rm ,*}}{\rm -}t_{y{\rm ,*}}\right\|{\rm =}\sqrt{{\left({\rm \ }t_{x{\rm ,1}}{\rm -}t_{y{\rm ,1}}\right)}^{{\rm 2}}{\rm +}{\left({\rm \ }t_{x{\rm ,2}}{\rm -}t_{y{\rm ,2}}\right)}^{{\rm 2}}{\rm +\dots +}{\left(t_{x,N{\rm -}{\rm 1}}{\rm -}t_{y,N{\rm -}{\rm 1}}\right)}^{{\rm 2}}{\rm +}{\left({\rm \ }t_{x,N}{\rm -}t_{y,N}\right)}^{{\rm 2}}}
\end{equation}
\\
Let's now use polyharmonic RBF with $\varepsilon = 1$:
\begin{equation}
d_{x,y}{\rm =\ }\varphi {\rm (}\left\|t_{x{\rm ,*}}{\rm -}t_{y{\rm ,*}}\right\|{\rm )=}\left\|t_{x{\rm ,*}}{\rm -}t_{y{\rm ,*}}\right\|
\end{equation}
\\
\textbf{Remark.} Since we are considering the norm then $d_{x,y} = d_{y,x}$. Follows the symmetry of $D$.
\\
Next, we define $W$ which is a $K{\rm \times }M$ matrix:
\begin{align}
W{\rm =}D^{{\rm -}{\rm 1}}S{\rm =}\left[ \begin{array}{ccc}
 \begin{array}{c}
w_{{\rm 1,1}} \\ 
 \begin{array}{c}
w_{{\rm 2,1}} \\ 
\vdots  \\ 
w_{k{\rm -}{\rm 1,1}} \end{array}
 \\ 
w_{k{\rm ,1}} \end{array}
 &  \begin{array}{ccc}
 \begin{array}{c}
w_{{\rm 1,2}} \\ 
 \begin{array}{c}
w_{{\rm 2,2}} \\ 
\vdots  \\ 
w_{k{\rm -}{\rm 1,2}} \end{array}
 \\ 
w_{k{\rm ,2}} \end{array}
 &  \begin{array}{c}
{\rm \dots } \\ 
 \begin{array}{c}
 \\ 
\ddots  \\ 
 \end{array}
 \\ 
{\rm \dots } \end{array}
 &  \begin{array}{c}
w_{{\rm 1,}M{\rm -}{\rm 1}} \\ 
 \begin{array}{c}
w_{{\rm 2,}M{\rm -}{\rm 1}} \\ 
\vdots  \\ 
w_{k{\rm -}{\rm 1,}M{\rm -}{\rm 1}} \end{array}
 \\ 
w_{k,M{\rm -}{\rm 1}} \end{array}
 \end{array}
 &  \begin{array}{c}
w_{{\rm 1,}M} \\ 
 \begin{array}{c}
w_{{\rm 2,}M} \\ 
\vdots  \\ 
w_{k{\rm -}{\rm 1,}M} \end{array}
 \\ 
w_{k,M} \end{array}
 \end{array}
\right]
\end{align}
\\
The current position matrix $L$ can be expressed as:
\begin{equation}
L{\rm =[} \begin{array}{ccc}
 \begin{array}{cc}
{\rm \ }l_{{\rm 1}} & l_{{\rm 2}} \end{array}
 & {\rm \dots } &  \begin{array}{cc}
l_{N{\rm -}{\rm 1}} & l_N \end{array}
 \end{array}
{\rm ]}
\end{equation}
\\
And finally:
\begin{equation}
\xi {\rm =[} \begin{array}{ccc}
 \begin{array}{cc}
{\rm \ }\varphi {\rm (}\left\|L{\rm -}t_{{\rm 1,*}}\right\|{\rm )} & \varphi {\rm (}\left\|L{\rm -}t_{{\rm 2,*}}\right\|{\rm )} \end{array}
 & {\rm \dots } &  \begin{array}{cc}
\varphi {\rm (}\left\|L{\rm -}t_{K{\rm -}{\rm 1,*}}\right\|{\rm )} & \varphi {\rm (}\left\|L{\rm -}t_{K{\rm ,*}}\right\|{\rm )} \end{array}
 \end{array}
{\rm ]} 
\end{equation}
\\
So we are now able to compute:
\begin{equation}
\Omega {\rm =}\xi W{\rm =[} \begin{array}{ccc}
 \begin{array}{cc}
{\rm \ }{\Omega }_{{\rm 1}} & {\Omega }_{{\rm 2}} \end{array}
 & {\rm \dots } &  \begin{array}{cc}
{\Omega }_{M{\rm -}{\rm 1}} & {\Omega }_M \end{array}
 \end{array}
{\rm ]}
\end{equation}
\\
\textbf{Remark}. If the current position is equal to the k-row of $T$, then $\Omega $ will be equal to the k-row of $S$.
\\
This solver has a lot of applications, not only in computer graphics to drive outputs but also, for instance, in Neural networks.

\subsection{\textit{The pseudocode}}

\begin{lstlisting}[language=python, frame=single]
Samples_Positions = LIST()  
Sample_Values = LIST()  
	  
Matrix_T = Create_matrix_from_list(Samples_Positions)  
Matrix_S = Create_matrix_from_list(Sample_Values)  
  
Epsilon = INT()  
  
Matrix_D = Compute_Distance_Matrix(Matrix_T)  
Kernel = Create_Kernel(SQUARED(Matrix_D), Epsilon)  
Inverted_kernel = Kernel.Interpolate(INTERPOLATION_TYPE).Invert()  
Weights_Matrix = Inverted_kernel * Matrix_S  
Current_Position = MATRIX()  
Current_Distance_Matrix = Compute_Distance_Matrix(Current_Position)  
  
Result = Current_Distance_Matrix * Weights_Matrix  
\end{lstlisting}

\section{Extending the RBF solver to quaternions}

\subsection{\textit{On quaternion blending}}
Being able to interpolate multiple quaternions, blending them or finding an average is not an easy task. $Slerp$ can't be extended in a straight forward way to multiple Quaternions due to the non-commutativity of quaternion product. 
\\
Over the years a few different approaches have been developed. An interesting paper from NASA \cite{Ma Ch Cr Os}, for instance, shows a solid mathematical model that is used for estimating the attitude (orientation) of a star.
\\
However we will use a different approach. Our goal is to blend Quaternions while still being able to operate with matrices since the blending depends on the computation of the RBF Solver. 
\\
As we saw in Section 3 Paragraph 1 , Quaternions generate a Lie Algebra and their Tangent Space is ${{\mathbb R}}^{{\rm 3}}$. The idea then is to map the Quaternions to be vectors in ${{\mathbb R}}^{{\rm 3}}$ with the $log$ function and work with them as ordinary vectors, proceed with the computation of the solver and finally convert them back to a Quaternion with the $exp$ function. 
\\
\\
\textbf{Definition}:
\begin{equation}
\overline{q}{\rm =}q_e{\exp  \left(\sum^K_{i{\rm =1}}{w_i{\ln  {\rm (}q_eq_i{\rm )}\ }}\right)\ }{\rm =}q_ee^{\sum^K_{i{\rm =1}}{w_i{\ln  {\rm (}q_eq_i{\rm )}\ }}}
\end{equation}
\\
Where $\overline{q}$ is the mean/blended Quaternion, $q_e$ is the multiplicative identity (or the base orientation), $w_i$ is the vector-weight associated to the i-th Quaternion $q_i$.

\subsection{\textit{Implementing the algorithm}}

\begin{lstlisting}[language=python, frame=single]
Samples_Positions = LIST()  
Sample_Values = LIST()  
	  
Matrix_T = Create_matrix_from_list(Samples_Positions)  
Matrix_S = Create_matrix_from_list(Sample_Values)  
  
Epsilon = INT()  
  
Matrix_D = Compute_Distance_Matrix(Matrix_T)  
Kernel = Create_Kernel(SQUARED(Matrix_D), Epsilon)  
Inverted_kernel = Kernel.Interpolate(INTERPOLATION_TYPE).Invert()  
LOG_Quats_Matrix = MATRIX(Matrix_S.Rowscount(), Matrix_S.Colscount())  
Quaternion_ID = Make_Identity_Quat()  
  
FOR i IN Matrix_S.Rowscount():  
    Sample_Quat = QUATERNION(Matrix_S[i][0],  
                             Matrix_S[i][1],  
                             Matrix_S[i][2]  
                             Matrix_S[i][3])  
    Sample_Quat_LOG = (Quaternion_ID * Sample_Quat.Normalized()).Log()  
    LOG_Quats_Matrix[i][0] = Sample_Quat_LOG.x  
    LOG_Quats_Matrix[i][1] = Sample_Quat_LOG.y  
    LOG_Quats_Matrix[i][2] = Sample_Quat_LOG.z  
    LOG_Quats_Matrix[i][3] = Sample_Quat_LOG.w  
	  
Weights_Matrix = Inverted_kernel * Matrix_S  
Current_Position = MATRIX()  
Current_Distance_Matrix = Compute_Distance_Matrix(Current_Position)  
	  
Linear_Result = Current_Distance_Matrix * Weights_Matrix  
Result = MATRIX(Linear_Result.Rowscount(), Linear_Result.Colscount())  
	  
FOR i IN Linear_Result.Rowscount():  
    LOG_Quat = QUATERNION(Linear_Result[i][0],  
                          Linear_Result[i][1],  
                          Linear_Result[i][2]  
                          Linear_Result[i][3])  
    EXP_Quat = LOG_Quat.EXP()  
    Result[i][0] = EXP_Quat.x  
    Result[i][1] = EXP_Quat.y  
    Result[i][2] = EXP_Quat.z  
    Result[i][3] = EXP_Quat.w  



\end{lstlisting}

\section{A practical use}
All of this has been used to create a plugin in C++ for Autodesk Maya \cite{Gou}. 
\\
For a full explanation of the development process the interested reader is remanded to \cite{Dol}.
\\ 
The intent is to show one possible way of how such a tool could be used down the pipeline in a working environment to improve the quality or speed up the process. 
\\
For this it was used Maya 2018 and Visual Studio on a machine running Windows 10.

\subsection{\textit{Correcting scapula orientation with the RBF}}

In this paragraph let's rig a realistic clavicle/scapula using the quaternion RBF solver. First thing let's import the model and let's place a few joints. 
\begin{center}
\includegraphics[width=62.8mm, height=46.8mm]{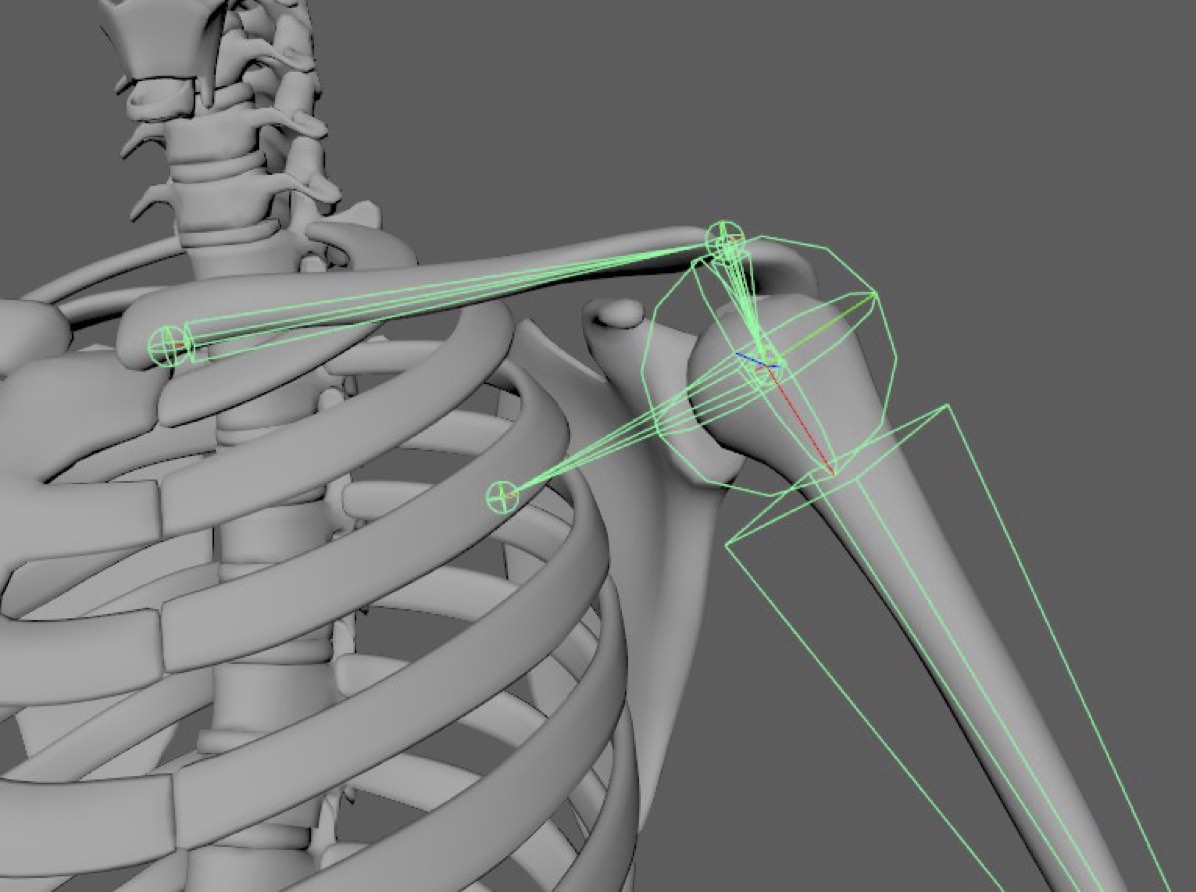}

Figure 13 - Anatomy model and joints placement.
\end{center}
We can now create a simple-chain (no rotate plane) IK Handle from the clavicle to the shoulder. This will enable us to simply move the handle and get the clavicle to rotate accordingly. The scapula will look broken, it might intersect the ribcage or aiming at the wrong direction. We can simply fix it by rotating the scapula joint. Once we are happy with how we fixed it let's duplicate the scapula joint. 
\begin{center}
\includegraphics[width=63.6mm, height=31.8mm]{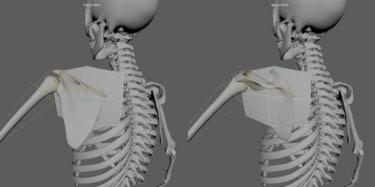}

Figure 14 - Clavicle front up, comparison scapula before (left) and after (right) fix.
\end{center}
If we iterate the same process for all 8 + 1 (the neutral position), after duplicating out the joints for each pose we should have something like this:
\begin{center}
\includegraphics[width=63.2mm, height=48.8mm]{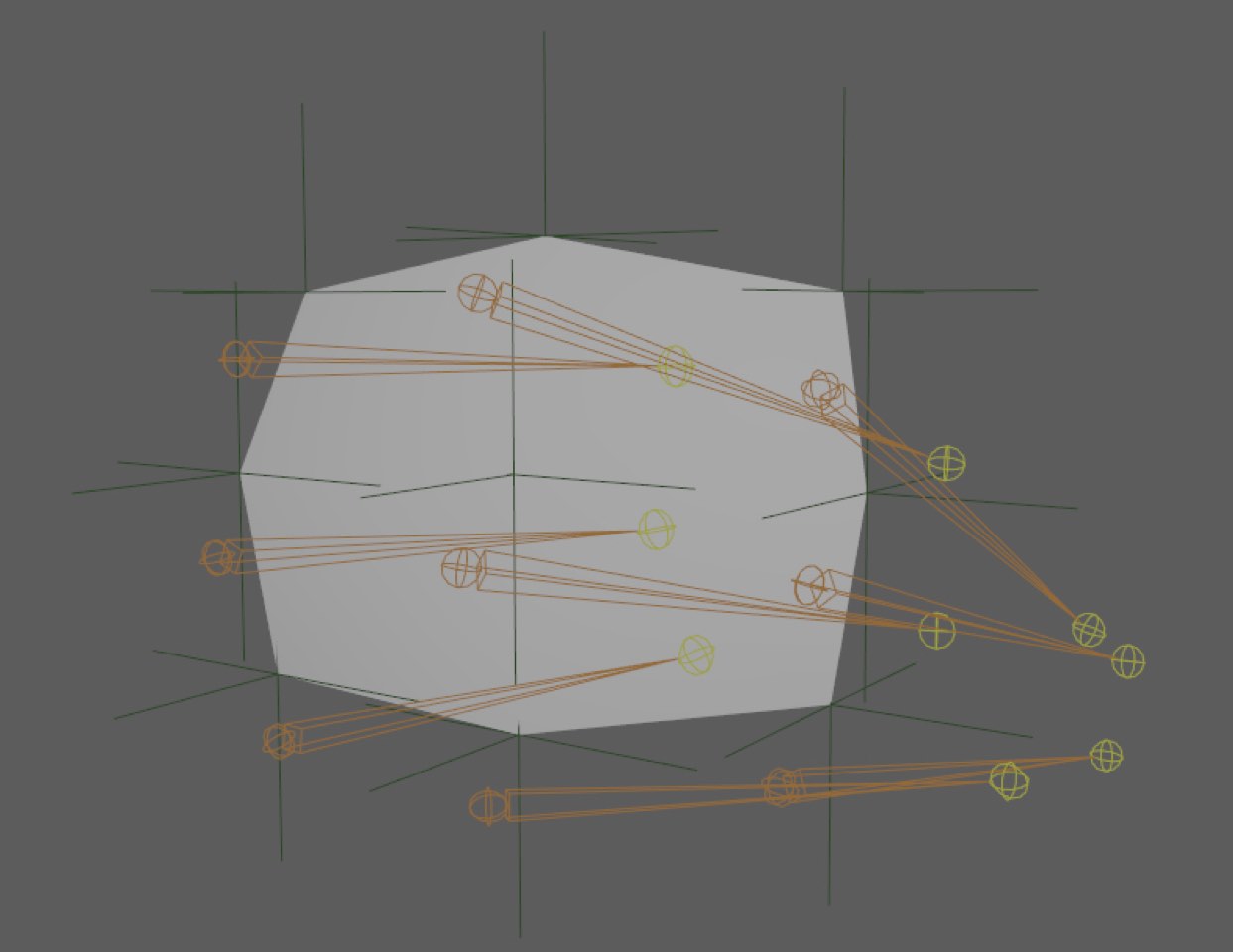}

Figure 15 - All clavicle and scapula coded positions.
\end{center}
The green crosses (called locators) visible in Figure 15 represent the positions of the clavicle-end, while the joints represent the right orientation of the scapula for each position of the clavicle. Before creating the solver and making the connections, let's clean it up and use simple cubes in place of the joints. 

\begin{center}
\includegraphics[width=63.0mm, height=49.6mm]{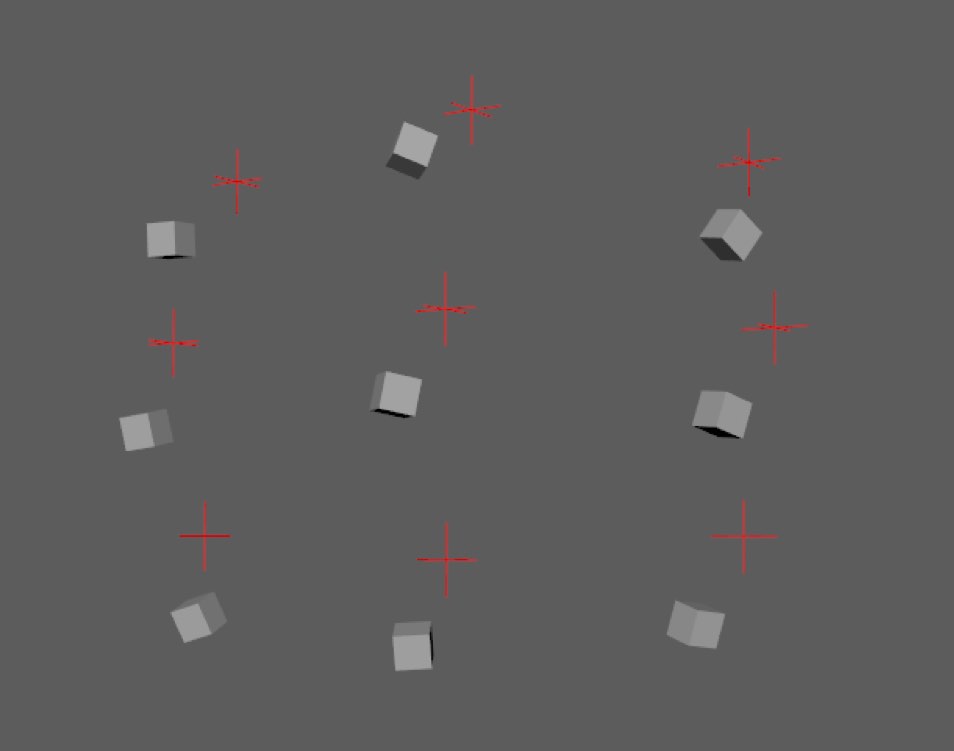}

Figure 16 - Scapula samples cleaned.
\end{center}
Let's create the ddQRbf (name chosen for the plugin developed) node and do the necessary connections. The output of the solver will go as input in a node that converts Quaternions to euler rotation and the output of this node will go into the scapula joint. The node network is shown below:
\begin{center}
\includegraphics[width=87.0mm, height=46.0mm]{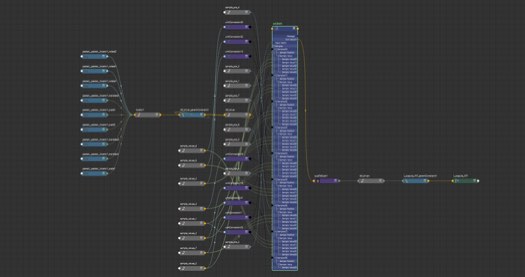}

Figure 17 - Scapula QRbf rig network.
\end{center}
The last thing we are left to do is to compare, for each sample, how it looks with and without the solver. In the following screenshots the red scapula is without the solver, the blue one is with the solver. 

\begin{center}
\includegraphics[width=100mm, height=100mm,]{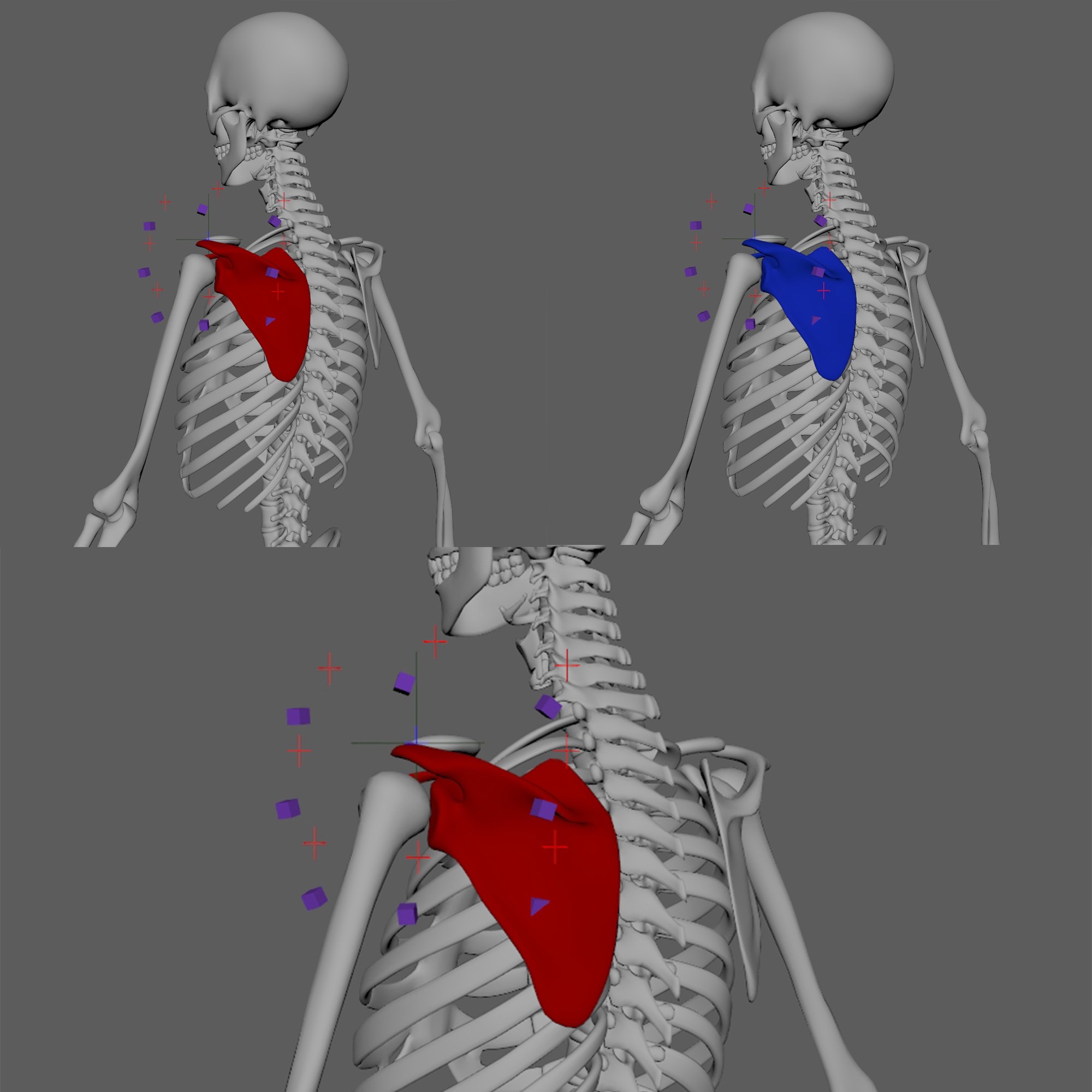}

Figure 18 -- Clavicle/Scapula, neutral sample.
\end{center}

\begin{center}
\includegraphics[width=100mm, height=100mm]{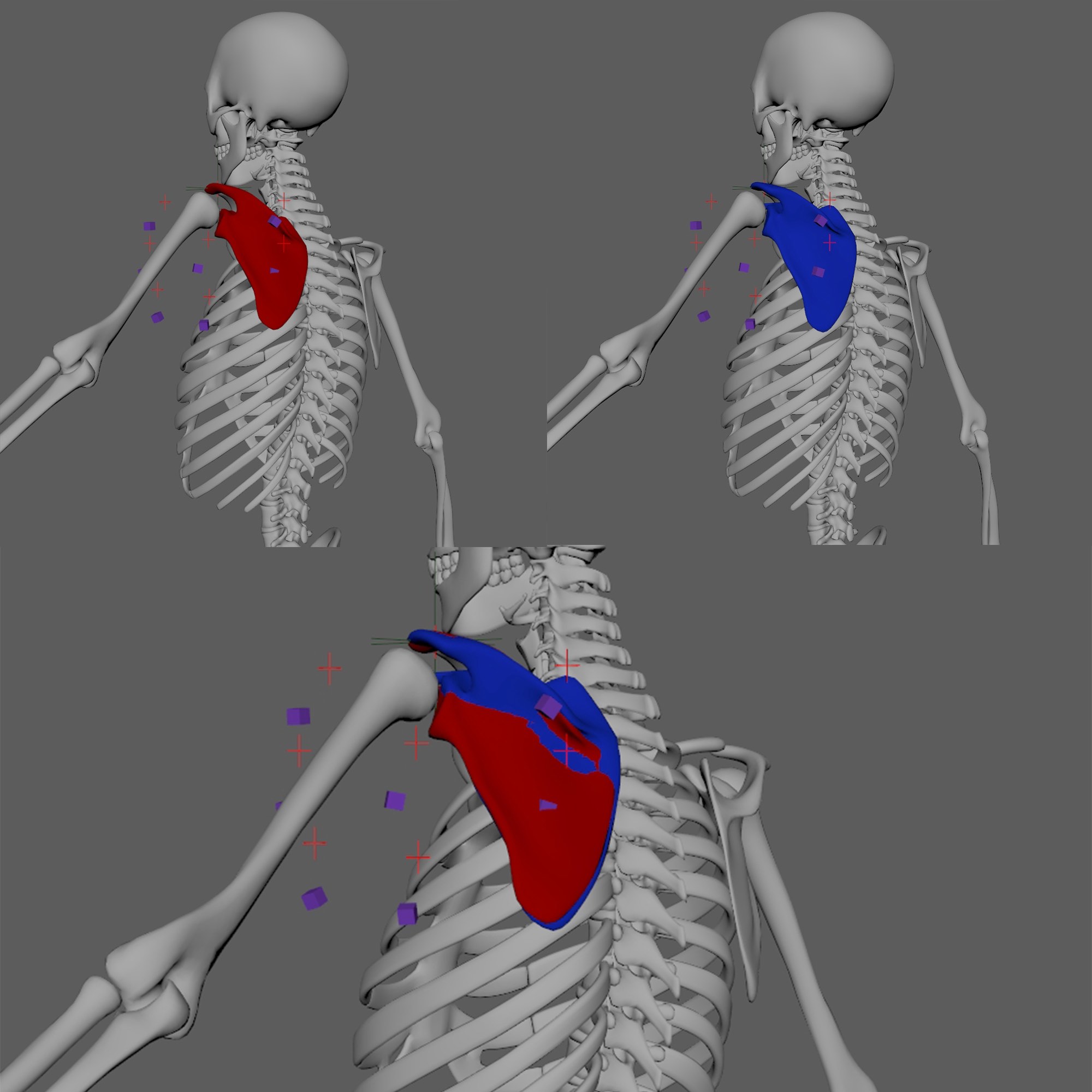}

Figure 19- Clavicle/Scapula, MidUp sample.
\end{center}

\begin{center}
\includegraphics[width=100mm, height=100mm]{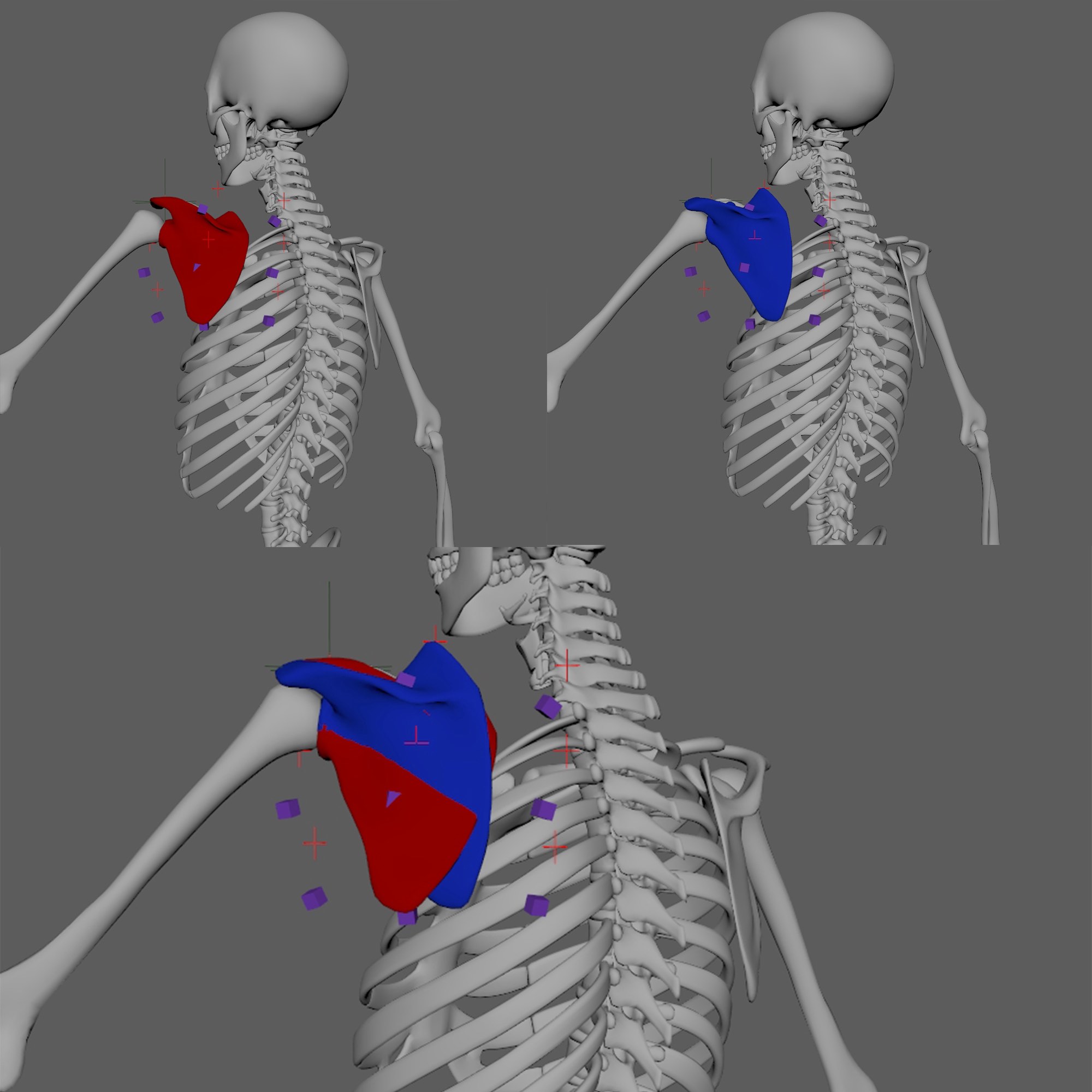}

Figure 20 - Clavicle/Scapula, FrontUp sample.
\end{center}

\begin{center}
\includegraphics[width=100mm, height=100mm]{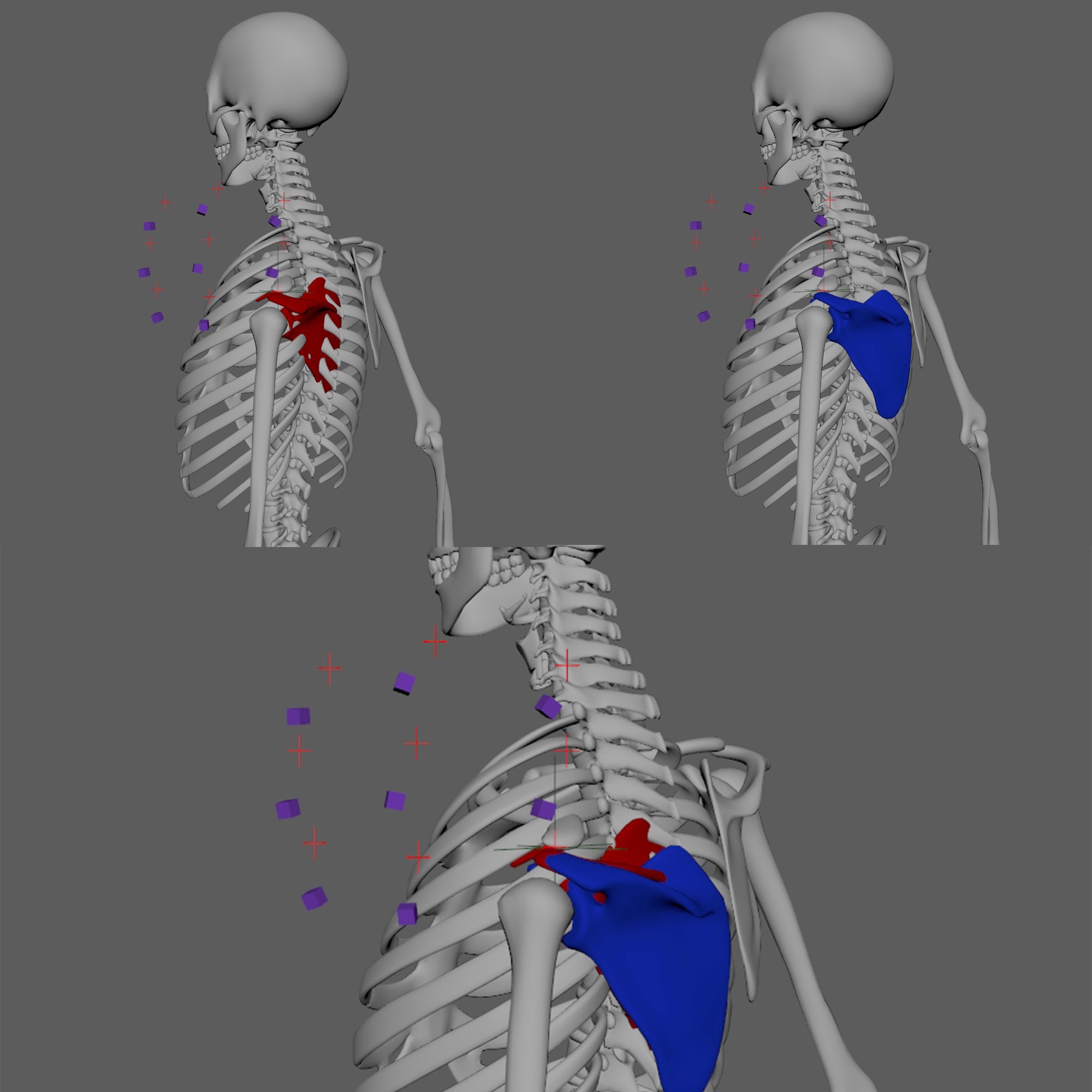}

Figure 21 - Clavicle/Scapula, BackDown sample.
\end{center}

It is obvious the great improvement on all of those poses: Intersections are fixed, the rotations are more anatomical and simulate how the muscles pull the scapula and the interpolation is very stable and organic. 
\\
\\
\textbf{Remark.} At first this might seem like a trivial result which is potentially achievable by using other type of techniques. However, let's not forget that this algorithm, applied in this plugin, is establishing a $new$ way of doing things. Furthermore this is just one of the infinite way that the Quaternions RBF solver could be used: indeed a classical RBF working with Euler angles will more likely encounter problems and surely be affected by the infamous $Gimbal$ $Lock$ at some point.

\section{Conclusions}
The main result, at least of mathematical interest, is related to the weighted blend of Quaternions, necessary to create the RBF solver. Using abstract algebra notions, such as Lie Algebra, Exponential Map and Tangent Space, it was possible to derive a formula that would map Quaternions to their Tangent Space, namely \textit{${{\mathbb R}}^3$}, which allowed to treat them just as simple vectors. More precisely, the idea was to map the Quaternions to be vectors in \textit{${{\mathbb R}}^3$} with the \textit{$log$} function and work with them as ordinary vectors, proceed with the computation of the solver and finally convert the result back to a Quaternion with the \textit{$exp$} function. In mathematical terms\textit{:}
\\
\[\overline{q}=q_e{\exp  \left(\sum^K_{i=1}{w_i{\ln  (q_eq_i)\ }}\right)\ }=q_ee^{\sum^K_{i=1}{w_i{\ln  (q_eq_i)\ }}}\] 
\\
Where $\overline{q}$ is the mean/blended Quaternion, $q_e$ is the multiplicative identity (or the base orientation), $w_i$ is the vector-weight associated to the \textit{i}-th Quaternion $q_i$.
\\
Finally, in Section 6, when got to the software development, this formula had to be adapted to work with arrays of Quaternions and arrays of weights. This process was challenging but eventually, once figured out how to compute the weights, there were not any more surprises. 
\\
As consequence of the that formula, the second main result is having the RBF Solver properly working and compiled. This could be used in a number of different ways down the production. In this paper it was shown a possible usage where the RBF Solver is used as a way to properly control and automate the orientation of the scapula according to the motion of the clavicle. Even though there are alternative ways to automate the scapula, this combo with Quaternions + RBF solver is a solid approach that proved to have a lot of potential. 
\\
A possible future line of research could be summarized as following:
\\
\begin{enumerate}
\item  \textit{Mathematical direction;} the available knowledge on transformations, both in two and three dimensions, is quite extensive. Especially with the fast growth of industries such as videogames, computer graphics, robotics and aeronautics the need for reliable and easy-to-implement mathematical models drove a lot of attention to the topic. However, these problems are normally solved from an engineering point of view which doesn't leave much space to an algebraic perspective. In this sense, exploring transformations is an already done, and almost closed, job, but figuring out new method and models that relies both on engineering and more elegant (abstract) algebraic methods is still a huge field to explore.  The formula for multiple Quaternions blending used in our RBF Solver is actually a great example of this. 

\item  \textit{Software direction;} the most straightforward action that can be taken is to improve the current solver by further optimizing the algorithm. Researches can be conducted to write other types of solvers. For example, it might be possible to write an IK (Inverse Kinematics) Quaternions-compatible solution or, perhaps, researches might be conducted towards deformers based on Quaternions. One big application of Quaternions might be related to dynamics simulation for both rigid and soft bodies.
\end{enumerate}

\section{Acknowledgements}

D. Dolci wishes to thank all of Animschool fellows, especially the Head of Character Program David Gallagher. Further he wishes to thank Yuri for his friendship throughout the years. He was a great inspiration to pursue this career.

\end{document}